\documentclass{article}

\usepackage{graphicx}
\usepackage{latexsym}
\usepackage{amssymb}

\usepackage{amssymb,cite} 

\def \BEA { \begin{eqnarray}}
\def \EEA {\end{eqnarray}}
\def \BE {\begin{equation}}
\def \EE {\end{equation}}
\def\d{\mathrm{d}}

\def \WDS #1 {\mbox{$\Phi_{#1}^{S}$}}
\def \WDA #1 {\mbox{$\Phi_{#1}^{A}$}}
\def \WD #1 {\mbox{$\Phi_{#1}$}}

\def \bl {\mbox{\boldmath{$\ell$}}}

\def \bn {\mbox{\boldmath{$n$}}}
\def \bk {\mbox{\boldmath{$k$}}}
\def \hbk {\mbox{\boldmath{$\hat k$}}}
\def \hbn {\mbox{\boldmath{$\hat n$}}}
\def \hbm #1 {\mbox{\boldmath{$\hat m^{(#1)}$}}}
\def \bm {\mbox{\boldmath{$m$}}}
\def \bk {\mbox{\boldmath{$k$}}}

\newcommand{\be}{\begin{equation}}
\newcommand{\ee}{\end{equation}}
\newcommand{\beqn}{\begin{eqnarray}}
\newcommand{\eeqn}{\end{eqnarray}}
\newcommand{\pa}{\partial}
\newcommand{\ba}{\begin{array}}
\newcommand{\ea}{\end{array}}
\newcommand{\pp}{{\it pp\,}-}

\def \BEAH {\begin{eqnarray*}}
\def \EEAH {\end{eqnarray*}}
\def \BEA {\begin{eqnarray}}
\def \EEA {\end{eqnarray}}
\def \BDM {\begin{displaymath}}
\def \EDM {\end{displaymath}}

\def \pull {{{{\frac{1}{2}}}}}

\def \S {S}

\newcommand{\M}[3] {{\stackrel{#1}{M}}_{{#2}{#3}}}
\newcommand{\m}[3] {\!\!{\stackrel{\hspace{.3cm}#1 \ 0}{m}}_{\!\!\!{#2}{#3}}\,}

\def \kS {{\cal H}}
\def \KS {{KS }}
\def \KSS {{KS}}

\newtheorem{proposition}{Proposition}

\begin{document}

\title{Higher dimensional Kerr-Schild spacetimes}

\author{Marcello Ortaggio\thanks{ortaggio(at)math(dot)cas(dot)cz}, Vojt\v ech Pravda\thanks{pravda@math.cas.cz}, Alena Pravdov\' a\thanks{pravdova@math.cas.cz} \\
Institute of Mathematics, Academy of Sciences of the Czech Republic \\ \v Zitn\' a 25, 115 67 Prague 1, Czech Republic}
\date{\today}

\maketitle

\abstract{We investigate general properties of Kerr-Schild (KS) metrics in $n>4$ spacetime dimensions. First, we show 
that the Weyl tensor is of type II or more special if the null \KS vector $\bk$ is geodetic (or, equivalently, if  $T_{ab}k^ak^b=0$). 
We subsequently specialize to vacuum \KS solutions, which naturally split into two families of non-expanding and expanding metrics. 
After demonstrating that {\em non-expanding} solutions are equivalent to the known class of vacuum Kundt solutions of Weyl type N, we analyze {\em expanding} solutions in detail. We show that they can only be of the type II or D, and we characterize optical properties of the multiple 
Weyl aligned null direction (WAND) $\bk$. In general, $\bk$ has caustics corresponding to curvature singularities. In addition, it is generically shearing. Nevertheless, we arrive at a possible `weak' $n>4$ extension of the Goldberg-Sachs theorem, {limited to the \KS class, which matches} previous conclusions for general type III/N solutions. 
In passing, properties of Myers-Perry black holes and black rings 
related to our results are also briefly discussed.}

\section{Introduction}

Kerr-Schild spacetimes \cite{KerSch652} have played an important role in four-dimensional general relativity. In particular, all vacuum \KS solutions have been known for some time \cite{KerSch652,DebKerSch69,Urbantke72}. They are a subset of algebraically special spacetimes, and the corresponding \KS null congruence is a geodetic, shearfree, repeated principal null direction of the Weyl tensor. Notably, they include the Kerr metric, arguably one of the physically most important known exact solutions of Einstein's equations in vacuum.
Other well-known vacuum solutions which can be put in the \KS form are, for example, pp-waves and type N Kundt waves. In addition, the \KS class can also admit (aligned) electromagnetic or matter fields, so as to include, for instance, the Kerr-Newmann metric, the Vaidya solution and \pp waves coupled to a null Maxwell field or to pure radiation. (We refer the reader to \cite{Stephanibook} for a comprehensive review and for a number of original references.)

In recent years, gravity in more than four spacetime dimensions has become an active area of ongoing studies, mainly motivated by the increasing interest in string theory and extra-dimensional scenarios. In this context, the \KS ansatz led to the remarkable discovery of rotating vacuum black holes in an arbitrary dimension $n>4$ \cite{MyePer86}. However, several features of higher-dimensional gravity have proven to be substantially different from the four-dimensional case, and more insight would be desirable. In this paper we explore systematically general properties of \KS metrics in higher dimensions. The main focus will be on Einstein's theory in vacuum, bearing in mind, however, that some of our results may have applications also in different contexts.

In addition to the intrinsic interest for possible specific \KS solutions, an advantage of the \KS ansatz is its mathematical tractability, {which enables one to study it analytically in full generality}. We thus present it also to illustrate a concrete application of the recently developed higher-dimensional classification of the Weyl tensor \cite{Coleyetal04,Milsonetal05}
and Newman-Penrose formalism \cite{Pravdaetal04,OrtPraPra07} (cf~\cite{PraPra08} for another very recent application). This will also enable us to present a partial extension (restricted to \KS solutions) of the Goldberg-Sachs theorem \cite{GolSac62,NP} to higher dimensions, thus making contact with the previous results of \cite{Pravdaetal04}. 
 The plan of the paper is as follows.

In Sec.~\ref{sec_general} we consider general \KS spacetimes and study constraints on the compatible form of the energy-momentum tensor. In particular, we show that the null \KS vector $\bk$ is geodetic if and only if $T_{ab}k^ak^b=0$ (similarly as in four dimensions \cite{Stephanibook}).

In Sec.~\ref{sec_geodetic} we specialize to  the case with a geodetic \KS vector $\bk$ and show that in that case the spacetime is of Weyl type II or more special. 

In the rest of the paper we further restrict ourselves to vacuum solutions. These naturally split into a non-expanding and an expanding class.
In Sec.~\ref{Sec_nonexp} we study the non-expanding class and show that it is equivalent to type N Kundt vacuum solutions. As a consequence, we also observe that $n>4$ pp-waves cannot  always be cast in the \KS form, as opposed to the four-dimensional case. 

The more complex expanding case is analyzed in Sec.~\ref{Sec_exp}. From a subset of the vacuum equations we obtain an {\em optical constraint}, i.e. a purely geometric condition on the KS null congruence  $\bk$ in the flat background. This is then used to show that expanding vacuum KS spacetimes are necessarily of Weyl type II or D. By integration of the Ricci identities we subsequently determine the dependence of the Ricci rotation coefficients (in particular, of optical quantities) on an affine parameter $r$ along the \KS congruence. This enables us also to find the $r$-dependence of the \KS function and thus to study general basic properties of \KS geometries. In particular, we discuss a `weak' extension of the Goldberg-Sachs theorem valid for vacuum \KS spacetimes, which are generically shearing. We also demonstrate, in the `generic' case, the presence of a curvature singularity at a caustic of $\bk$. This is in particular true for all expanding non-twisting spacetimes, on which we comment as a special subset of \KS solutions.

We conclude in Sec.~\ref{sec_conclusions} with a summary and final remarks. In appendix~\ref{app_Riemann} frame components of the Riemann tensor are given for KS spacetimes with a geodetic  $\bk$. Subsequent appendices~\ref{app_special}--\ref{app_ricci} contain several proofs and technical details related to the results presented in the main text.

\subsection{Preliminaries}

We use the notation of \cite{Pravdaetal04} (see also \cite{OrtPraPra07}) and, in particular, we introduce a null frame $\bm^{(0)} =\bn$, $\bm^{(1)} =\bl$, $\bm^{(i)}$ ($i=2,\dots, n-1$). 
The only non-zero scalar products are $\ell^a n_a = 1$ and $m^{(i)a}m^{(j)}_a=\delta_{ij}$ ($a,b=0,1,\ldots,n-1$), and the spacetime metric reads $g_{a b} = 2\ell_{(a}n_{b)} + \delta_{ij} m^{(i)}_a m^{(j)}_b$ (sum over $i,j$). Directional derivatives along frame vectors 
will be denoted by $D \equiv \ell^a \nabla_a$, $\bigtriangleup  \equiv n^a \nabla_a$, $\delta_i \equiv m^{(i)a} \nabla_a$. The full set of the corresponding Ricci rotation coefficients is defined in \cite{Pravdaetal04,OrtPraPra07}. In particular, we will often use the definition $\ell_{a;b}=L_{cd}m^{(c)}_a m^{(d)}_b$ (sum over $c,d$). In fact, it will be convenient to adapt the frame vector $\bl$ to coincide with the \KS null vector $\bk$. When $\bk$ is geodetic, we will denote by $r$ the corresponding affine parameter. Quantities that do not depend on $r$ will be denoted by a subscript or superscript index $0$ (e.g., $\kS_0$, $s_{(i)}^0$, etc.).

Let us also anticipate here that we will be using five types of indices with different ranges, namely
\beqn
 & &a,b,c,\ldots=0,1,\ldots,n-1 , \qquad i,j,k,\ldots=2,\dots, n-1 , \nonumber \\
 & & \alpha,\beta,\ldots=2p+2,\ldots,m+1 , \qquad \rho,\sigma,\ldots=m+2,\ldots,n-1 , \\
 & & \mu,\nu=1,\ldots,p , \qquad \mbox{with } \ 0\le2p\le m\le n-2, \nonumber
\eeqn
where $p$ and $m$ are fixed integers defined later on. Einstein's summation convention is employed except for indices $\mu,\nu$ (for which summation will be indicated explicitly), or when at least one of the repeated indices is in brackets (e.g., there will be no summation over $i$ in $DA_{ij}=-2s_{(i)}A_{ij}$), unless (only in very few exceptional cases) stated otherwise. Note that for indices $i,j,\ldots$ we will not distinguish between subscripts and superscripts since there is no difference between covariant and contravariant components.

When referring to equations presented in previous papers it will be sometimes convenient to denote them by the corresponding equation number followed by the reference number, e.g. eq.~(11f,\cite{OrtPraPra07}).

\section{General case}
\label{sec_general}

In this section we define \KS spacetimes and we study  their general properties, without imposing Einstein's equations and without requiring the \KS null congruence to be geodetic.

\subsection{Ansatz and geodetic condition}

We study an $n$-dimensional spacetime with a metric in the \KS form
\BE
g_{ab} = \eta_{ab} - 2 \kS k_a k_b ,  \label{Ksmetr}
\EE
with $\eta_{ab}$ being the Minkowski metric diag$(-1,1,....,1)$,  $\kS$ a scalar function
and $k_a$ a 1-form that is assumed to be null with respect to  $\eta_{ab}$, i.e. $\eta^{ab}k_a k_b=0$ ($\eta^{ab}$ is defined as the inverse of $\eta_{ab}$ and $\eta^{ab}\neq g^{ac}g^{bd}\eta_{cd}$). From (\ref{Ksmetr}) it follows for the corresponding  vector that $k^a \equiv \eta^{ab} k_b= g^{ab} k_b$, so that $k^a$ is  null also with  respect to
$g_{ab}$, and that the inverse metric has the form
\BE
g^{ab} = \eta^{ab} + 2 \kS k^a k^b . \label{Ksmetrinv}
\EE
One also gets for the metric determinant
\be
 g=\eta=-1 .
 \label{determinant}
\ee

Straightforwardly from the definition of the Christoffel symbols we obtain 
\be
 \Gamma^{a}_{\ bc}=-(\kS k^ak_b)_{,c}-(\kS k^ak_c)_{,b}+\eta^{ad}(\kS k_bk_c)_{,d}+2\kS k^ak^d(\kS k_bk_c)_{,d} ,
 \label{Christoffel}
\ee
from which
\BE
\Gamma^{a}_{\ bc} k^b k^c =0, \qquad \ \ \Gamma^{a}_{\ bc} k_a k^b =0, \qquad \ \ \Gamma^{a}_{\ ab} =  0 ,
\EE
and consequently also
\BE
k_{a;b} k^b = k_{a,b} k^b, \qquad \ \ k^{a}_{\ ;b} k^b = k^{a}_{\ ,b} k^b.
 \label{k_der}
\EE
In particular, this implies that {\em $k^a$ is geodetic in the flat geometry $\eta_{ab}$ iff it is geodetic in the full geometry $g_{ab}$.}
We also have
\BE
\Gamma^{a}_{\ bc} k_a = (\kS k_b k_c)_{;d} k^d, \qquad \Gamma^{a}_{\ bc} k^b = -(\kS k^a k_c)_{;d} k^d ,
\label{Gamma_k}
\EE
\BE
k_{a;c}k^a=k_a,_c k^a=0,\qquad  k_{a;bc}k^a+k_{a;b}k^a \ _{;c}=k_a,_{bc}k^a+k_a,_b k^a,_c=0 .
\EE

Since $\Gamma^{a}_{\ ab} =  0$, we can express the Ricci tensor as
\BE
R_{ab} = \Gamma^c_{\ ab,c} - \Gamma^{c}_{\ db} \Gamma^{d}_{\ ac} .
\EE
After employing the above formulas, Einstein's equations for the projected component $R_{ab} k^a k^b$ read 
\BE
R_{ab} k^a k^b = 2 \kS g^{ab}(k_{a;c} k^c) ( k_{b;d} k^d) = \kappa_0 T_{ab} k^a k^b.
 \label{T00}
\EE

From now on, when using a null frame $\bl$, ${\bn}$, $\bm^{(i)}$, we will make the convenient choice 
\be
 \bl=\bk ,
\ee 
adapted to the \KS ansatz. 

Using eq.~(\ref{T00}) and (see \cite{Pravdaetal04})
\BE
k_{a;b}k^b=L_{10}k_a+L_{i0}m^{(i)}_a\label{k_abkb}
\EE
{(sum over $i$)}, we can  formulate the following
\begin{proposition}
\label{prop_geod}
 The null vector $\bk$ in the \KS metric (\ref{Ksmetr}) of an arbitrary dimension is geodetic if and only if the energy-momentum
tensor satisfies ${T_{ab} k^a k^b=0}$.
\end{proposition}

Then, using frame indices, the condition of proposition~\ref{prop_geod} reads $T_{00}=0$ {(or, equivalently, $R_{00}=0$)}, and it holds iff the null frame components of $T_{ab}$ do not include a term proportional to 
$n_a n_b$. This condition is of course satisfied in the case of {\em vacuum} spacetimes, also with a possible cosmological constant, or in the presence of matter fields aligned  with the \KS vector $\bk$, such as an aligned Maxwell field (defined by $F_{ab} k^a  \sim k_b $)
or aligned pure radiation (i.e., $T_{ab} \sim k_a k_b $).

\subsection{Optics of the \KS vector field}

\label{subsec_optics}

It is interesting to discuss how the optical properties of $k_a$ in the two geometries $g_{ab}$ and $\eta_{ab}$ are related, i.e. to compare the Ricci rotation coefficients \cite{Pravdaetal04,OrtPraPra07} $L_{ab}$ and $\tilde L_{ab}$ defined with respect to $g_{ab}$ and $\eta_{ab}$, respectively. In order to do so one has to set up `null' frames for both metrics. 
Let $\bk$, ${\bn}$, $\bm^{(i)}$ be the frame for the full spacetime metric $g_{a b}$ as discussed above.
Then, from eq.~(\ref{Ksmetr}) it follows that $\eta_{a b} = 2k_{(a}[n_{b)}+\kS k_{b)}] + \delta_{ij} m^{(i)}_a m^{(j)}_b$ (sum over $i,j$). Thus, $\bk$, ${\mbox{\boldmath{$\tilde n$}}}$, $\bm^{(i)}$ is now a convenient frame for the flat metric $\eta_{ab}$, provided one takes
\be
 \tilde n_a=n_a+\kS k_a .
 \label{frames}
\ee

Now, from eqs.~(\ref{Gamma_k}) and (\ref{k_abkb}) one gets
\be
 k_{a;b}=k_{a,b}-(D\kS+2\kS L_{10})k_ak_b-2\kS L_{i0}m^{(i)}_{(a}k_{b)} 
 \label{covder}
\ee
{(sum over $i$)}.
Expanding $k_{a;b}$ on the frame $\bk$, ${\bn}$, $\bm^{(i)}$ and $k_{a,b}$ on the frame $\bk$, ${\mbox{\boldmath{$\tilde n$}}}$, $\bm^{(i)}$ in terms of $L_{ab}$ and $\tilde L_{ab}$, respectively, and using~(\ref{frames}) one finds that $L_{ab}=\tilde L_{ab}$ except for $L_{11}=\tilde L_{11}-D\kS-\kS\tilde L_{10}$ and 
$L_{1i}=\tilde L_{1i}-\kS\tilde L_{i0}$.
In particular, one has $L_{i0}=\tilde L_{i0}$ (so that $\bk$ is geodetic with respect to $g_{a b}$ iff it is geodetic with respect to $\eta_{a b}$, as already observed in section~\ref{sec_general}), and the matrix  
\be
 L_{ij}\equiv k_{a;b}m^{(i)a}m^{(j)b}=k_{a,b}m^{(i)a}m^{(j)b} 
\ee
is the same with respect to both metrics. The optical scalars shear, twist and expansion \cite{Pravdaetal04,OrtPraPra07} are thus unchanged\footnote{Cf~\cite{Xanthopoulos83} in the case $n=4$.} (note that when $\bk$ is geodetic this statement is independent of the particular choice (\ref{frames}) since $L_{ij}$ is then invariant under null rotations with $\bk$ fixed \cite{OrtPraPra07}).

\section{The case of a geodetic \KS vector field}

\label{sec_geodetic}

In the rest of the paper we will assume that the null \KS vector $\bk$ is {\em geodetic} and {\em affinely parametrized} (i.e., $L_{i0}=0=L_{10}$), which seems to be the simplest and physically most interesting case (recall proposition~\ref{prop_geod}). The following definitions \cite{Pravdaetal04,OrtPraPra07} will thus be useful:
\beqn
 & & S_{ij}\equiv L_{(ij)}=\sigma_{ij}+\theta\delta_{ij}, \qquad A_{ij}\equiv L_{[ij]} , \nonumber \label{optics} \\
 & &  \theta\equiv\textstyle{\frac{1}{n-2}}S_{ii} , \qquad \sigma^2\equiv\sigma_{ij}\sigma_{ij}, \qquad \omega^2\equiv A_{ij}A_{ij} .
\eeqn
We shall refer to $S_{ij}$, $\sigma_{ij}$ and $A_{ij}$ as the {\em expansion}, {\em shear} and {\em twist} matrices, respectively, and to $\theta$, $\sigma$ and $\omega$ as the corresponding scalars.

Under the geodetic assumption, eq.~(\ref{covder}) reduces to 
\be 
 k_{a;b}=k_{a,b}-(D\kS) k_a k_b ,
 \label{covder_geod}
\ee
which will be employed in the following. Using $k^{a}_{\ ,b}=\eta^{ac}k_{c,b}$ (and, from now on, denoting $(\ ),_d\eta^{cd}$ by $  (\ )^{,c}$, e.g. $k_a^{\ ,b}=\eta^{bc}k_{a,c}$ etc.)
 one easily finds $k_{a,b}k^{a,c}=k_{a;b}k^{a;c}$, $k_{a,b}k^{c,a}=k_{a;b}k^{c;a}$, $k_a^{\ ,b}k^a_{\ ,c}=k_a^{\ ;b}k^a_{\ ;c}$, $k_a,_{bc}k^b=-k_a,_b k^b,_c$,  $k_a,_{bc}k^b k ^c=0$ and similar identities, also to be used in following calculations. In addition, certain expressions will take a more compact form if we write them in terms of the Ricci rotation coefficients and of directional derivatives as 
\beqn
 & & k^{a}_{\ ,a}=L_{ii}=S_{ii} , \qquad k_{a,b}k^{a,b}=L_{ij}L_{ij}, \qquad k_{a,b}k^{b,a}=L_{ij}L_{ji}, \nonumber \\   & & \kS_{,a}k^a=D\kS, \qquad  \kS_{,ab}k^ak^b=D^2\kS .
\eeqn

From eq.~(\ref{Christoffel}) we thus get 
\BEA
& & \Gamma^{e}_{\ ab,e} =
(\kS k_a k_b)_{, \ \ e}^{\ e} - (\kS k_a k^e)_{,be} -(\kS k_b k^e)_{,ae}  \nonumber \\
& & \qquad\qquad {}+ 
2\left[ \kS  D^2\kS 
+  (D\kS  
+ \kS L_{ii} 
) D\kS 
\right] k_a k_b , \\
& & \Gamma^{e}_{\ fb} \Gamma^{f}_{\ ae}=2\left[  (D\kS
)^2 -  2\kS^2 \omega^2 
\right] k_a k_b,
\EEA

The Ricci tensor then reads\footnote{This formula for $R_{ab}$ is equivalent to eq.~(32.10) given in \cite{Stephanibook} for $n=4$, up to rewriting partial derivatives as covariant ones.}
\BEA
R_{ab} &=& (\kS k_a k_b)_{, \ \ e}^{\ e} - (\kS k_a k^e)_{,be} -(\kS k_b k^e)_{,ae}  \label{Ricci} \nonumber \\
&&{}+ 2\kS\left[   D^2\kS
+   L_{ii} 
D\kS
 +2\kS \omega^2
 \right] k_a k_b  .
\EEA
Consequently, { $\bk$ is an eigenvector of the Ricci tensor, i.e.} 
\BE
R_{ab} k^b =-[D^2\kS+(n-2)\theta D\kS+2\kS\omega^2]k_a. \label{eigenRicci}
\EE

For certain applications it may be useful to observe that the mixed Ricci components, i.e. $R^a_{\ b}=(\kS k^a k_b)_{, \ \ s}^{\ s} - (\kS k^a k^s)_{,bs} -(\kS k_b k^s)_{, \ \ s}^{\ a}$, are linear in $\kS$ \cite{GurGur75,DerGur86}. The Ricci scalar is thus also linear and reads
\be 
 \hspace{-.12cm} R=-2\left[D^2\kS+2(n-2)\theta D\kS+\kS(n-2)(n-3)\theta^2+\kS(\omega^2-\sigma^2)\right] .
\ee

So far we have worked in Minkowski coordinates adapted to the flat background metric $\eta_{ij}$. Note, however, that expressions for scalars, such as frame components of tensors, do not depend on that choice.

\subsection{Frame components of the Ricci tensor}

From eq.~(\ref{Ricci}), one finds the non-vanishing frame components of the Ricci tensor
\beqn
 R_{01} & = & -[D^2\kS+(n-2)\theta D\kS+2\kS\omega^2] , \label{R01} \\
 R_{ij} & = & 2\kS L_{ik}L_{jk}-2[D\kS+(n-2)\theta\kS]S_{ij} , \label{Rij} \\
 R_{11} & = & \delta_i(\delta_i \kS)+(N_{ii}-2\kS L_{ii})D\kS+(4L_{1j}-2L_{j1}-\M{i}{j}{i})\delta_j\kS-L_{ii}\Delta\kS \nonumber \\
 & & {}+2\kS\big(2\delta_iL_{[1i]}+4L_{1i}L_{[1i]}+L_{i1}L_{i1}-L_{11}L_{ii} \nonumber \\ 
 & & {}+2L_{[1j]}\M{j}{i}{i}-2A_{ij}N_{ij}-2\kS\omega^2\big) , \label{R11} \\
 R_{1i} & = & -\delta_i(D\kS)+2L_{[i1]}D\kS+2L_{ij}\delta_j\kS-L_{jj}\delta_i\kS \nonumber \\ & & \hspace{-1cm}  {}+2\kS\big(\delta_jA_{ij}+A_{ij}\M{j}{k}{k}-A_{kj}\M{i}{k}{j}-L_{jj}L_{1i}+3L_{ij}L_{[1j]}+L_{ji}L_{(1j)}\big) . \label{R1i}   
\eeqn
Note that $R_{00}=0=R_{0i}$ identically. 

\subsection{Algebraic type of the Weyl tensor}

The full set of the frame components of the Riemann tensor is given in appendix~\ref{app_Riemann} for any \KS geometry with a geodetic, affinely parametrized \KS vector. Thanks to $R_{0i0j}=R_{010i}=R_{0ijk}=0$ and to $R_{00}=R_{0i}=0$, for the Weyl tensor we find immediately
\be
 C_{0i0j}=0 , \qquad C_{010i}=0, \qquad C_{0ijk}=0 .
\ee
The Weyl tensor components with boost order~2 and 1 are thus identically zero, which enables us to conclude  
\begin{proposition}
 \label{prop_II}
 Kerr-Schild spacetimes (\ref{Ksmetr}) in arbitrary dimension with a geodetic  \KS vector $\bk$ are of Weyl type II (or more special), with $\bk$ being a WAND  of order of alignment $\geq$ 1.   
\end{proposition}

Note that the \KS null vector $\bk$ must indeed be geodetic for a wide class of matter fields, in particular in vacuum (cf proposition~\ref{prop_geod}), so that proposition~\ref{prop_II} applies in those cases.

Another interesting result follows from the observation \cite{PraPraOrt07} that spacetimes (not necessarily of the \KS class) which are either static or stationary with `expansion' and `reflection symmetry' can be only of the Weyl types G, I$_i$, D or O (see \cite{PraPraOrt07} for details and precise definitions). Taking the `intersection' of this family with the set of \KS spacetimes considered in proposition~\ref{prop_II}, we can conclude that in arbitrary dimension $n\ge4$ 
\begin{proposition}
\label{prop_stationary}
 All static spacetimes of the \KS class with a geodetic $\bk$ are of Weyl type D or O. All stationary spacetimes of the \KS class with `reflection symmetry' and with a geodetic, expanding $\bk$ are of Weyl type D or O. In both cases, if the Weyl tensor is non-zero, $\bk$ is a multiple WAND.
\end{proposition} 

As an immediate consequence, the higher-dimensional rotating black holes of Myers and Perry (indeed obtained using the \KS ansatz \cite{MyePer86}), are necessarily of Weyl type D in any dimension, as we anticipated in \cite{PraPraOrt07} (this had  previously been demonstrated by an explicit calculation of the Weyl tensor in \cite{Hamamotoetal06}). This also applies to uniform black strings, since adding flat dimensions to a \KS metric clearly preserves the \KS structure. On the other hand, proposition~\ref{prop_II} implies that five-dimensional vacuum black rings \cite{EmpRea02PRL} do not admit a \KS representation, since they are of Weyl type I$_i$ \cite{PraPra05}. 
Let us also emphasize that proposition~\ref{prop_stationary} is not restricted to vacuum \KS solutions. For instance, it can also be used to conclude that static black holes with electric charge (and, possibly, a cosmological constant) \cite{Tangherlini63} are also of Weyl type D.


\subsection{Vacuum equations}

Vacuum solutions must satisfy $R_{ab}=0$. We note from eqs.~(\ref{R01})--(\ref{R1i}) that the Ricci tensor components $R_{01}$ and $R_{ij}$ are simple and do not involve Ricci rotation coefficients other than $L_{ij}$, which characterize the optical properties of $\bk$. It is thus natural to start from the corresponding vacuum equations. 
Imposing $R_{ij}=0$ gives\footnote{Hereafter, for brevity we shall write $\kS^{-1}D\kS=D\ln\kS$, where it is understood that $\ln\kS$ should be replaced by $\ln|\kS|$ whenever $\kS<0$.}
\be
 (D\ln\kS)S_{ij}=L_{ik}L_{jk}-(n-2)\theta\S_{ij} .
 \label{vac_ij}
\ee
Contracting eq.~(\ref{vac_ij}) with $\delta^{ij}$ we obtain
\beqn
 (n-2)\theta (D\ln\kS) & = & L_{ik}L_{ik}-(n-2)^2\theta^2 \nonumber \\
 & = & \sigma^2+\omega^2-(n-2)(n-3)\theta^2, 
 \label{Rii=0}
\eeqn 
while its tracefree part (i.e., $R_{ij}-\frac{R_{kk}}{n-2}\delta_{ij}=0$) is\footnote{For $n=4$ this reduces to $(D\ln\kS)\sigma_{ij}=2\sigma_{k(i}A_{j)k}$.} 
\beqn
(D\ln\kS)\sigma_{ij}=\left({\sigma^2}_{ij}-\textstyle{\frac{1}{n-2}}\sigma^2\delta_{ij}\right)-\left({A^2}_{ij}+\textstyle{\frac{1}{n-2}}\omega^2\delta_{ij}\right)  \nonumber \label{tracefree}\\ 
{}+2\sigma_{k(i}A_{j)k}-(n-4)\theta\sigma_{ij} .
\eeqn

Next, the equation $R_{01}=0$ requires
\be
 D^2\kS+(n-2)\theta D\kS+2\kS\omega^2=0 .
 \label{R01=0}
\ee
{Using eq.~(\ref{Rii=0}), this can also be rewritten as $D^2\kS=[-\sigma^2-3\omega^2+(n-2)(n-3)\theta^2]\kS$.}

Note that eq.~(\ref{Rii=0}) involves the function $\kS$ in a non-trivial way only when $\theta\neq0$. It will thus be convenient to study non-expanding and expanding solutions separately in the following sections. The remaining vacuum equations seem to be of little help in a general study, and there is no need to write them down at this stage.


\section{Non-expanding vacuum solutions}

\label{Sec_nonexp}

\subsection{Einstein's equations}

The case $\theta=0$ turns out to be somewhat special, since from eq.~(\ref{Rii=0}) we immediately get $\sigma=0=\omega$,  
and thus 
\be 
 L_{ij}=0 , \label{Kundt_Rii}
\ee 
and from this and eq.~(\ref{R01=0})
\be
 D^2\kS=0 , \label{Kundt_R01}
\ee  
i.e., $\kS=\kS_0+{\cal G}_0 r$ (recall that in our notation $r$ denotes an affine parameter along $\bk$ and $D\kS_0=0=D{\cal G}_0$). The vacuum equations $R_{ij}=0$ and $R_{01}=0$ are thus identically satisfied. Note, in particular, that the sole non-expanding condition implies that we are restricted to a subset of the higher-dimensional {\em Kundt class} of non-expanding, non-shearing and non-twisting vacuum solutions. 

Setting $L_{ij}=0$ in the Ricci components (\ref{R11}) and (\ref{R1i}), the remaining vacuum equations read
\beqn
 & & \delta_i(\delta_i \kS)+N_{ii}D\kS+(4L_{1j}-2L_{j1}-\M{i}{j}{i})\delta_j\kS \nonumber \\ & & \qquad {}+2\kS\big(2\delta_iL_{[1i]}+4L_{1i}L_{[1i]}+L_{i1}L_{i1}+2L_{[1j]}\M{j}{i}{i}\big)=0 , \label{Kundt_R11} \\
 & & \delta_i(D\kS)-2L_{[i1]}D\kS=0 . \label{Kundt_R1i}   
\eeqn

\subsection{Equivalence with Kundt solutions of type N}

\subsubsection{Non-expanding \KS solutions are of type N}

We have seen above that vacuum non-expanding \KS solutions are Kundt spacetimes. In general, the higher-dimensional Kundt class is known to be of Weyl type II or more special, provided $R_{00}=0=R_{0i}$ (thus in particular in vacuum) \cite{OrtPraPra07}. However, here we show that, in arbitrary dimension, vacuum \KS spacetimes of the Kundt class are restricted to  type N. 
In fact, it has  already been shown in section~\ref{sec_geodetic} that the components of the Weyl tensor with boost weight~2 and 1 vanish. Using eqs.~(\ref{Kundt_Rii}), (\ref{Kundt_R01}) and (\ref{Kundt_R1i}) together with eqs.~(\ref{R0101})--(\ref{R1ijk}) we also find that all Weyl components with boost weight~0 and $-1$ are zero (since in vacuum $R_{abcd}=C_{abcd}$). The Weyl tensor is thus of type N, q.e.d..

\subsubsection{Kundt solutions of type N are \KS}

We have thus demonstrated that vacuum solutions of the \KS class with a non-expanding \KS vector are a subset of Kundt spacetimes of Weyl type N. We now show that the converse is also true, namely that vacuum Kundt solutions of Weyl type N are of the \KS form, so that the two families of solutions coincide.
  
Vacuum Kundt solutions of Weyl type N belong to the family of spacetimes with vanishing scalar invariants (VSI) \cite{Coleyetal04vsi,Coleyetal06}. We can thus begin with the higher-dimensional {VSI} metric \cite{Coleyetal03,Coleyetal06}
\BE
 \d s^2=2\d u\left[\d
r+H(r,u,x^k)\d u+W_{i}(r,u,x^k)\d x^i\right]+\delta_{ij}\d x^i\d
x^j , \label{Kundt}
\EE
with $i,j,k=2,\dots,n-1$, $k_a\d x^a = \d u$ being the multiple WAND. 

 Similarly as for $n=4$, higher-dimensional vacuum Kundt spacetimes of Weyl type N consist of two invariantly defined subfamilies \cite{Coleyetal03,Coleyetal06}: Kundt waves (with $L_{1i}=L_{i1}\neq 0$) and \pp waves (for which $L_{1i}=L_{i1}=0=L_{11}$). It is convenient to discuss these subfamilies separately.

The~metric functions $H$ and $W_{i}$ for higher-dimensional type N Kundt waves are \cite{Coleyetal06} 
\BEA
W_{2} & = & -2\frac{r}{x^{2}}, \nonumber \\
W_{s} & = & x^{q}B_{qs}(u)+C_{s}(u),  \label{KundtN}\\
H & = & \frac{r^2}{2(x^{2})^{2}}+H^0(u,x^{i}), \nonumber 
\EEA
 where $s,q=3,\dots,n-1$, $B_{qs}=-B_{sq}$ { and $H^{0}$ must obey a field equation given in \cite{Coleyetal06}.} 
As shown in \cite{Coleyetal06}, metric (\ref{Kundt}), (\ref{KundtN}) is flat for
\BE
H^{0}(u,x^{i})=H^{0}_{\mbox{\tiny flat}} =
\frac{1}{2} \sum_{s=3}^{n-1} W_s^2 +x^2 F_0(u)+x^2 x^iF_{i}(u),
\EE
where $F_0(u)$ and $F_i(u)$ are arbitrary functions of $u$.  
Therefore metric (\ref{Kundt}), (\ref{KundtN}) is in the \KS form for an arbitrary choice of $H^{0}(u,x^{i})$, since it can always be rewritten as ${\rm d}s^2={\rm d}s^2_{\mbox{\tiny flat}}+(H^{0}-H^{0}_{\mbox{\tiny flat}}){\rm d}u^2$.

A similar argument can be used to show that also higher-dimensional type N \pp waves belong to the \KS class -- the corresponding metric functions $W_{i}$ and $H^{0}_{\mbox{\tiny flat}}$ entering (\ref{Kundt}) can be found in \cite{Coleyetal06}.\footnote{There is a typo in eq.~(96) of \cite{Coleyetal06}: just drop the inequality $m\le n$ in the second sum. We thank Nicos Pelavas for correspondence on this point.}

We can thus summarize the results of this section in the following (see~theorem~32.6 of \cite{Stephanibook} for $n=4$)
\begin{proposition}
 In arbitrary dimension $n\ge 4$, the Kerr-Schild vacuum spacetimes with a non-expanding \KS congruence $\bk$ coincide with the class of vacuum Kundt solutions of Weyl type N.
\label{prop_nonexp}
\end{proposition}

\subsubsection{ \KS does not include all \pp waves for $n>4$}

A comment on some differences between $n=4$ and $n>4$ dimensional \pp waves is now in order. 
{By a natural extension of the $n=4$ terminology of \cite{Stephanibook},} in any dimension \pp waves are defined as spacetimes (not necessarily of the \KS form) admitting a covariantly constant null vector $\bl$, i.e. $\ell_{a;b}=0$. It then follows directly from the definition of the Riemann tensor that
\BE
R_{abcd} \ell^d = 0. \label{typeNnecc}
\EE

For four-dimensional vacuum spacetimes this is equivalent to the definition of the type N \cite{Stephanibook}. It is also known that, in addition to being of Weyl type N, all four-dimensional vacuum \pp waves can be cast in the \KS form \cite{Stephanibook}. 

By contrast, in higher dimensions eq.~(\ref{typeNnecc}) is only a necessary, but not sufficient condition for type N. 
{It only says that the type is II (or more special) \cite{PraPra05}, and in fact for $n>4$ one can easily construct explicit vacuum \pp waves of  types III, II and D, as we now briefly discuss}. 

Vacuum \pp wave metrics of Weyl type III can be obtained directly by specializing results of \cite{Coleyetal06} to the vacuum case. One simple five-dimensional example is metric (\ref{Kundt}) with
\BEA
 & & W_2=0, \quad W_3=h(u) x^2 x^4,  \quad W_4 = h(u) x^2 x^3,\\
 & & H=H_0 = h(u)^2 \left[ \frac{1}{24} \left( (x^3)^4+ (x^4)^4 \right)+ h^0( x^2,x^3,x^4) \right],
\EEA
where $h^0(x^2,x^3,x^4)$ is linear in $x^2$, $x^3$, $x^4$. 

In addition, one can also construct $(n_1+n_2)$-dimensional vacuum \pp waves of Weyl type II, {e.g.} by taking a direct product of a $n_1$-dimensional vacuum 
\pp wave of Weyl type N or III with an Euclidean $n_2$-dimensional Ricci-flat (but non-flat) metric (with both $n_1,n_2\ge4$, and $n_1\ge5$ if the $n_1$-dimensional \pp wave is of type III).\footnote{The proof of these statements is straightforward and we just sketch it. First, all the mentioned products are really \pp waves since a covariantly constant vector field $\bl$ defined in the $n_1$-dimensional geometry can be trivially lifted to the $(n_1+n_2)$-dimensional product geometry, in which it will still be covariantly constant. Such products are necessarily of Weyl type II or more special (with the lifted $\bl$ being a multiple WAND), as follows from the decomposable form of the Weyl tensor of product geometries \cite{Coleyetal04,PraPraOrt07}. They cannot be of Weyl type III or N since they inherit non-zero curvature invariants (e.g., Kretschmann) from the $n_2$-dimensional Euclidean space. They cannot be of type D since, starting from a frame adapted to the canonical form of the (type N or III) Weyl tensor of the $n_1$-dimensional spacetime, the components of negative boost weight turn out to be unchanged under null rotations, and thus cannot be set to zero. The only possible Weyl type is thus indeed II.} Similarly, $(n_1+n_2)$-dimensional vacuum \pp waves of Weyl type D arise if one takes a direct product of a $n_1$-dimensional flat spacetime with an Euclidean $n_2$-dimensional curved, Ricci-flat space (with $n_1\ge2$, $n_2\ge4$).

As a direct consequence of proposition~\ref{prop_nonexp}, such $n>4$ \pp waves of Weyl type II, D or III do not admit a \KS form. They thus represent a `counterexample' to both the $n=4$ results, namely they are (higher-dimensional) vacuum \pp waves that are neither of type N nor of the \KS form. 

Note, in addition, that \pp waves of Weyl type II and D cannot belong to the VSI class of spacetimes (which is compatible only with  types III, N, O \cite{Coleyetal04vsi}), and therefore they necessarily possess some non-vanishing curvature invariants. In table 1 we summarize the {aforementioned} properties
of various types of {vacuum} \pp waves {in four (4D) and higher dimensions (HD)}.\footnote{In fact, in view of our previous comments a similar table applies to all Kundt solutions, in which case type III appears also in 4D \cite{Stephanibook}, again VSI but not KS.}

\begin{table}[h]
\begin{center}
	\begin{tabular}{|c|c|c|c|}
	\hline
			 &&&\\[-3mm]
			 & \ Weyl type \  & \ \ KS \ \  & \ \ VSI \ \  \\
			 &&&\\[-3mm]
			 \hline &&&\\[-10pt]
   4D	& N  &  $\surd$  &  $\surd$ \\[1pt]
		\hline &&&\\[-10pt]
		HD & N & $\surd$ & $\surd$ \\[1pt]
		\hline &&&\\[-10pt]
		HD & III & X & $\surd$ \\[1pt]
		\hline &&&\\[-10pt]
		HD & \ II\,(D) \ & \ X \ & \ X \ \\[1pt]
		\hline 
		\end{tabular}
		\end{center}
		\caption{Properties
of various types of vacuum \pp waves.}
\end{table}

\section{Expanding vacuum solutions}
\label{Sec_exp}

Non-expanding vacuum solutions of the \KS class have been fully classified according to the discussion of the previous section. Let us now consider solutions with an expanding $\bk$, i.e. $\theta\neq 0$.

\subsection{Consequences of Einstein's equations}

For $\theta\neq 0$, from (\ref{Rii=0}) one can write $D\ln\kS$ as 
\be
 D\ln\kS=\frac{L_{ik}L_{ik}}{(n-2)\theta}-(n-2)\theta . 
 \label{DS}
\ee
Substituting into (\ref{vac_ij}) one gets
\be
 L_{ik}L_{jk}=\frac{L_{lk}L_{lk}}{(n-2)\theta}S_{ij} .
 \label{optical_const}
\ee
Interestingly, this equation is independent of the \KS function $\kS$ and it is thus a purely geometric condition on the \KS null congruence $\bk$ in the Minkowskian `background' $\eta_{ab}$ (recall the discussion of subsection~\ref{subsec_optics}). We will thus refer to it as the {\em optical constraint}.\footnote{This constraint is of course identically satisfied in the trivial case $\sigma=0=\omega$ (i.e., $L_{ij}=S_{ij}=\theta\delta_{ij}$) which includes, e.g., the Schwarzschild-Tangherlini solution \cite{Tangherlini63}. Less trivial examples are provided by the \KS congruence of static black strings, for which $\sigma\neq 0$ and $\omega=0$, or of rotating black strings and Myers-Perry black holes (which were indeed constructed in \cite{MyePer86} using the \KS anstatz), both having $\sigma\neq 0\neq\omega$. (To explicitly verify these statements one may use the corresponding optical quantities calculated in any dimension in \cite{Pravdaetal04,PraPraOrt07}.) 

On the other hand, a simple example of a geodetic null congruence {\em violating} the optical constraint is provided by the vector field $\bk=\sqrt{1+f^2(\phi)}\pa_t+f(\phi)\pa_\rho+\pa_z$ in the flat geometry $\d s^2=-\d t^2+\d\rho^2+\rho^2\d\phi^2+\d z^2+\delta_{AB}\d y^A\d y^B$ (with $A,B=1,\ldots,n-4$). This geodetic null congruence is expanding and shearing for $f\neq 0$, and violates the optical constraint if $f_{,\phi}\neq 0$, in which case it is also twisting.} Its consequences on the form of $L_{ij}$ will be studied in detail in appendix~\ref{app_constraint} and \ref{app_ricci}, and they will be employed in the following sections.

 Using~(\ref{optics}), eq.~(\ref{optical_const}) can also be rewritten in terms of the shear and twist matrices as
\beqn  (n-2)\theta\left[\left({\sigma^2}_{ij}-\textstyle{\frac{1}{n-2}}\sigma^2\delta_{ij}\right)-\left({A^2}_{ij}+\textstyle{\frac{1}{n-2}}\omega^2\delta_{ij}\right)+2\sigma_{k(i}A_{j)k}\right] \nonumber \\ 
 =\left[\sigma^2+\omega^2-(n-2)\theta^2\right]\sigma_{ij} .
 \label{matrix_const}
\eeqn
This traceless equation is in fact equivalent to (\ref{tracefree}), after using (\ref{DS}) (since the trace of eq.~(\ref{optical_const}) is obviously an identity).\footnote{In odd dimensions, it implies that one cannot have $\sigma_{ij}=0$ if $A_{ij}\neq 0$ (this was already known in a more general context \cite{OrtPraPra07}).} Contracting (\ref{matrix_const}) with $\sigma_{ij}$ one finds the scalar constraint
\be
\sigma^2\left[\sigma^2+\omega^2-(n-2)\theta^2\right]-(n-2)\theta\left[{\sigma^2}_{ij}\sigma_{ij}-\sigma_{ij}{A^2}_{ji}\right]=0 .
 \label{scalar_const}
\ee 

So far we have discussed consequences of the vacuum equation ${R_{ij}=0}$. Next, one has to make sure that eq.~(\ref{R01=0}) (i.e., $R_{01}=0$) is now compatible with (\ref{DS}). In fact, taking the $D$-derivative of (\ref{DS}) and using the scalar Sachs equations of \cite{OrtPraPra07} and eq.~(\ref{scalar_const}), one exactly recovers eq.~(\ref{R01=0}). This is thus automatically satisfied, provided the preceding equations hold. Of course, when looking for an explicit solution one should also solve the remaining Einstein equations, namely $R_{11}=0$ and $R_{1i}=0$. These are, however, too involved unless one makes some further assumptions (such as the presence of symmetries, etc.). In any case, they are not needed in the following discussion.

\subsection{Algebraic type}

From the general result of proposition~\ref{prop_II}, we already know that the Weyl tensor of any \KS metric with a geodetic $\bk$ is of type II or more special. Here we show that, in fact, the types III and N are not possible for vacuum expanding solutions. 

By {\em reductio ad absurdum}, let us thus assume that all components of the Riemann (Weyl) tensor with boost weigh zero, i.e. (\ref{R0101})--(\ref{Rijkl}), vanish. In particular, we can multiply $R_{0i1j}$ by $L_{lj}$. Using the optical constraint (\ref{optical_const}), the condition $R_{0i1j}L_{lj}=0$ gives 
\be
 S_{il}D\ln\kS=-2A_{ki}S_{kl} ,
 \label{AkiSkl}
\ee
where we have dropped an overall factor $L_{jk}L_{jk}=\sigma^2+\omega^2+(n-2)\theta^2>0$. Using as a `spatial' basis an eigenframe of $S_{ij}$, we have $S_{ij}=\mbox{diag}(s_{(2)},s_{(3)},s_{(4)},\ldots)$. At least one eigenvalue must be non-zero, say $s_{(2)}\neq 0$. Then from (\ref{AkiSkl}) with $i=l=2$ we get
\be 
 s_{(2)}D\ln\kS=-2s_{(2)}A_{22} ,
\ee 
which obviously can never be satisfied since $A_{22}=0$ (and since the case $D\kS=0$ is `forbidden', cf~appendix~\ref{app_special}). This contradiction completes the proof of
\begin{proposition}
 In arbitrary dimension $n\ge 4$, Kerr-Schild vacuum spacetimes with an expanding \KS congruence $\bk$ are of algebraic type II or D.
\label{prop_exp}
\end{proposition} 

For a similar result for $n=4$ cf~\cite{McIHic88,Stephanibook}. 

\subsection{Fixing the $r$-dependence}

As detailed in appendices~\ref{app_constraint} and \ref{app_ricci}, the optical constraint~(\ref{optical_const}) and the Sachs equations \cite{OrtPraPra07} imply that there exists an appropriate frame, satisfying $\M{i}{j}{0}=0$, such that the matrix $L_{ij}$ takes the block diagonal form
\beqn
 L_{ij}=\left(\begin {array}{cccc} \fbox{${\cal L}_{(1)}$} & & &  \\
  & \ddots & & \\ 
 & & \fbox{${\cal L}_{(p)}$} & \label{L_general} \\
 & & & \fbox{$\begin {array}{ccc} & & \\ \ \ & \tilde{\cal L} \ \ & \\ & & \end {array}$}
  \end {array}
 \right) . 
 \eeqn
The first $p$ blocks are $2\times 2$ and the last block $\tilde{\cal L}$ is a $(n-2-2p)\times(n-2-2p)$-dimensional diagonal matrix. They are given by 
\beqn
 & & {\cal L}_{(\mu)}=\left(\begin {array}{cc} s_{(2\mu)} & A_{2\mu,2\mu+1} \nonumber \\
 -A_{2\mu,2\mu+1} & s_{(2\mu)} 
\end {array}
 \right) \qquad (\mu=1,\ldots, p) , \\
  & & s_{(2\mu)}=\frac{r}{r^2+(a^0_{(2\mu)})^2} , \quad A_{2\mu,2\mu+1}=\frac{a^0_{(2\mu)}}{r^2+(a^0_{(2\mu)})^2} , \label{s_A} \\
  & &  \tilde{\cal L}=\frac{1}{r}\mbox{diag}(\underbrace{1,\ldots,1}_{(m-2p)},\underbrace{0,\ldots,0}_{(n-2-m)}) , 
 \label{diagonal}
\eeqn
with $0\le 2p\le m\le n-2$. The integer $m$ denotes the rank of $L_{ij}$, so that $L_{ij}$ is non-degenerate when $m=n-2$.

Accordingly, the optical scalars are given by
\beqn
 & & (n-2)\theta=2\sum_{\mu=1}^p\frac{r}{r^2+(a^0_{(2\mu)})^2}+\frac{m-2p}{r} , \label{exp_nondeg} \\
 & & \omega^2=2\sum_{\mu=1}^p\left(\frac{a^0_{(2\mu)}}{r^2+(a^0_{(2\mu)})^2}\right)^2 , \label{twi_nondeg} \\
 & & \sigma^2=2\sum_{\mu=1}^p\left(\frac{r}{r^2+(a^0_{(2\mu)})^2}-\theta\right)^2+(m-2p)\left(\frac{1}{r}-\theta\right)^2 \nonumber \\
 & & \qquad\qquad {}+(n-2-m)\theta^2 . \label{she_nondeg} 
\eeqn
Note, in particular, that  expansion $\theta$ indicates the presence of a caustic at $r=0$, except when $2p=m$ (with $m$ even). 
Note also that  shear is generically non-zero (the special case $\sigma=0$ will be discussed below). The twist is zero if and only if $p=0$ (or, equivalently, all $a^0_{(2\mu)}$ vanish).

One can similarly fix the $r$-dependence of all Ricci rotation coefficients  (at least with the additional `gauge' condition $N_{i0}=0$). Since this is not needed in the following discussion, corresponding results are relegated to appendix~\ref{app_ricci}.

Knowing the form of $L_{ij}$ enables us to solve the vacuum equation~(\ref{DS}), where now $L_{ik}L_{ik}=(n-2)\theta r^{-1}$. 
The $r$-dependence of $\kS$ is thus given by
\be
 \kS=\frac{\kS_0}{r^{m-2p-1}}\prod_{\mu=1}^p\frac{1}{r^2+(a^0_{(2\mu)})^2} .
 \label{Hgeneral2}
\ee
This includes in particular the solution of~(\ref{DS}) in the case when its rhs vanishes, i.e. $D\kS=0$, which happens for $m=1$ (implying $p=0$). This is, however, incompatible with the Bianchi identities \cite{Pravdaetal04} as explained in appendix~\ref{app_special}. Hence, in the following we shall restrict to
\be
 2\le m\le n-2 . \label{m>1}
\ee

For example, the $r$-dependence in the special case of Myers-Perry solutions is obtained by setting $m=n-2$ in~(\ref{Hgeneral2}) (note, however, that function~(\ref{Hgeneral2}) is more readily compared with that of \cite{MyePer86} using the parametrization~(6) of \cite{CheLuPop06} for the direction cosines). Black holes with only one non-zero spin \cite{MyePer86}, in particular, correspond to the further choice $p=1$, and static black holes to $p=0$ (see also subsection~\ref{subsec_special}).

In general, the asymptotic behavior of $\kS$ for $r\to\infty$ is given by
\be
 \kS=\frac{\kS_0}{r^{m-1}}+{\cal O}(r^{-m-1}) ,
\ee 
i.e., the function $\kS$ behaves as a Newtonian potential in $(m+1)$ space dimensions with $r$ as a radial coordinate.

\subsection{On the Goldberg-Sachs theorem}

Let us now use the above results to comment on a partial extension to higher dimensions (but restricted to \KS solution) of the Goldberg-Sachs theorem. In four dimensions, this is a well-known theorem stating that in a vacuum (non-flat) spacetime, a null congruence is geodetic and shearfree if and only if it is a multiple principal null direction of the Weyl tensor \cite{GolSac62,NP,Stephanibook}.

For $n=4$, the matrix $L_{ij}$ associated with a generic null vector $\bl$ \cite{Pravdaetal04} is $2\times 2$ and can be written in terms of the standard Newman-Penrose notation \cite{Stephanibook} as \cite{OrtPraPra07}\footnote{Note that, exceptionally in this subsection only, the complex shear $\sigma$ does not coincide with the real shear scalar defined in section~\ref{sec_geodetic} and used throughout the paper (although they are simply related), but it is the usual Newman-Penrose spin coefficient.}
\be
 L_{ij}=-\frac{1}{2}\left(\begin {array}{cc} (\rho+\bar\rho)+(\sigma+\bar\sigma) & -i(\rho-\bar\rho)+i(\sigma-\bar\sigma) \nonumber \\
 i(\rho-\bar\rho)+i(\sigma-\bar\sigma) & (\rho+\bar\rho)-(\sigma+\bar\sigma)
\end {array}
 \right) .
\ee

In vacuum, when $\bl$ is a multiple principal null direction of the Weyl tensor the Goldberg-Sachs theorem implies $\sigma=0$, so that $L_{ij}$ reduces to
\be
 L_{ij}=-\frac{1}{2}\left(\begin {array}{cc} \rho+\bar\rho & -i(\rho-\bar\rho) \nonumber \\
 i(\rho-\bar\rho) & \rho+\bar\rho 
\end {array}
 \right) .
 \label{L_4dim}
\ee

We have observed in previous sections that for vacuum \KS spacetimes the \KS vector $\bl=\bk$ is indeed a (geodetic) multiple WAND for any $n\ge 4$. Accordingly, in the special case $n=4$, eqs.~(\ref{L_general})--(\ref{diagonal}) must be of the shearfree form (\ref{L_4dim}). It is easy to verify that this is indeed the case, since for $n=4$ we necessarily have $m=2$ (cf~(\ref{m>1})) and therefore $L_{ij}$ either coincides with one of the blocks ${\cal L}_{(\mu)}$ (for $p=1$, i.e. $\rho-\bar\rho\neq0$) or is proportional to the two-dimensional identity matrix (for $p=0$, i.e.  $\rho-\bar\rho=0$). 

However, it has been pointed out \cite{MyePer86,FroSto03,Pravdaetal04,OrtPraPra07,PraPraOrt07} that the Goldberg-Sachs theorem cannot be extended to higher dimensions in the most direct formulation `multiple WANDs are geodetic and shearfree in vacuum'. Indeed, we have already noticed  that $\bk$ is generically shearing. Nevertheless, for any $n>4$ our result (\ref{L_general})--(\ref{diagonal}) can be viewed as a `weaker' higher-dimensional Goldberg-Sachs theorem, albeit {\em restricted to \KS spacetimes}: in a suitable basis, {\em the matrix $L_{ij}$ (associated with the geodetic, multiple WAND $\bk$) consists of $2\times2$ blocks {that are} `shearfree'}, {i.e.} reflecting the four-dimensional shearfree condition (\ref{L_4dim}) in various orthogonal 2-planes.\footnote{Let us recall that we have been focusing on the expanding case $\theta\neq 0$, since $\theta=0\Rightarrow L_{ij}=0$ for vacuum \KS spacetimes (see section~\ref{Sec_nonexp}).} The only `exceptions' to this of course arise in odd spacetime dimensions (there will be an isolated one-dimensional block) and in the degenerate case $\det L=0$, in which one may add an arbitrary number of zeros along the diagonal (most simply, by just taking a direct product with flat dimensions). This `generalized Goldberg-Sachs condition' can also be expressed in a basis-independent form simply as (dropping the matrix indices) 
\be
 [S,A]=0 , \qquad A^2=S^2-{\cal F}S ,
 \label{GolSac_inv}
\ee
where $S=(L+L^T)/2$ and $A=(L-L^T)/2$ are the symmetric and antisymmetric parts of $L$, and ${\cal F}$ can be fixed as in~(\ref{cal_F}) by taking the trace. {The first equation in~(\ref{GolSac_inv}) is clearly equivalent to the condition $[L,L^T]=0$, i.e. $L$ is a (real) {\em normal} matrix.} The second equation in~(\ref{GolSac_inv}) is just a consequence of the first one when $n=4$.

Let us also emphasize that the above `weak Goldberg-Sachs theorem' has, in fact, been proven in more generality (i.e., without the \KS assumption) for higher-dimensional vacuum spacetimes of Weyl type III and N: the WANDs of such spacetimes must be geodetic, and the associated matrix $L_{ij}$ is indeed of the form~(\ref{L_general}), with only one non-zero $2\times 2$ block and zeros elsewhere \cite{Pravdaetal04}.\footnote{More precisely, this has been proven for all $n>4$ solutions of Weyl type N, for $n>4$ non-twisting solutions of type III, and for all $n=5$ solutions of type III. For twisting type III solutions with $n>5$ an extra assumption on the Weyl tensor was necessary \cite{Pravdaetal04}.}

For a possible complete extension of the Goldberg-Sachs theorem, one should study the remaining possibilities, i.e. $n>4$ vacuum spacetimes of Weyl type II and D that do not belong to the \KS family. In addition, one should also analyze the inverse implication, that is, whether in vacuum the existence of a geodetic null congruence with an associated matrix $L_{ij}$ of the form~(\ref{L_general}) implies that the algebraic type is II or more special (this is already known in the simple Kundt case $L_{ij}=0$ \cite{OrtPraPra07}). Let us recall that the `geodetic part' of the $n>4$ Goldberg-Sachs theorem has been discussed in \cite{PraPraOrt07}.

It is worth mentioning that a $n>4$ extension of the Robinson theorem 
has been proven in even dimensions \cite{HugMas88} assuming a 
generalization of the shearfree condition (see also 
\cite{NurRob02,Trautman02a,Trautman02b,MasTag08}) different from ours. 
The relation to our work will be discussed elsewhere.

\subsection{Weyl tensor components of boost weight zero}

The $r$-dependence of the matrix $L_{ij}$ determines explicitly also the $r$-dependence of the Weyl components with boost weight~0 (cf~appendix \ref{app_Riemann}). In order to express these components compactly, it is convenient to introduce a $(n-2) \times (n-2)$ matrix  $\WD{ij} \equiv C_{0i1j}$ along with its symmetric and antisymmetric parts $\WDS{ij} \equiv\WD{(ij)} $ and $\WDA{ij} \equiv\WD{[ij]} $. {All} boost order zero components of the Weyl tensor are then determined in terms of the $(n-2)(n-3)/2$ independent components of $\WDA{ij} $ and the $(n-1)(n-2)^2(n-3)/12$ independent components of $C_{ijkl}$, since $\WDS{ij} = - \textstyle{\frac{1}{2}} C_{ikjk}$, $C_{01ij}=2\WDA{ij} $, $C_{0101}=\Phi\equiv\WD{ii} $ \cite{Pravdaetal04,PraPraOrt07}.

It follows from appendix~\ref{app_Riemann} that, using the adapted frame of appendix~\ref{app_constraint}, the matrix $\WD{ij} $ inherits the block structure of $L_{ij}$, with {the only non-zero components and trace given by 
\beqn
 & & \WD{2\mu,2\mu} =\WD{2\mu+1,2\mu+1} =-2\kS A_{2\mu,2\mu+1}^2-s_{(2\mu)}D\kS ,\label{PhiS} \\
 & & \WD{2\mu,2\mu+1} =\WDA{2\mu,2\mu+1} =-D(\kS A_{2\mu,2\mu+1}) , \label{PhiA} \\
 & & \WD{\alpha\beta} =-\frac{1}{r}\delta_{\alpha\beta}D\kS , \qquad \Phi=D^2\kS=-(n-2)\theta D\kS-2\kS\omega^2 . \label{Phi_alpha} 
\eeqn

Up to index permutations, the non-zero $C_{ijkl}$ components are 
\beqn
 & & C_{2\mu,2\mu+1,2\mu,2\mu+1}=2\kS(3A_{2\mu,2\mu+1}^2-s_{(2\mu)}^2) ,\label{Cijkl} \\
 & & C_{2\mu,2\mu+1,2\nu,2\nu+1}=2C_{2\mu,2\nu,2\mu+1,2\nu+1}=-2C_{2\mu,2\nu+1,2\mu+1,2\nu} \nonumber \\
 & & \qquad\qquad\qquad\qquad =4\kS A_{2\mu,2\mu+1}A_{2\nu,2\nu+1}  , \\
 & & C_{2\mu,2\nu,2\mu,2\nu}=C_{2\mu,2\nu+1,2\mu,2\nu+1}=-2\kS s_{(2\mu)}s_{(2\nu)}, \\
 & & C_{(\alpha)(i)(\alpha)(i)}=-2\kS s_{(i)}r^{-1} ,\label{Cijkl3}
\eeqn
where $\nu\neq\mu$. 

Let us observe that all the above components fall off at $r\to\infty$ as $1/r^{m+1}$ or faster (in the non-twisting case, i.e. when all $A_{2\mu,2\mu+1}$ vanish, only $1/r^{m+1}$ terms are present,  cf~also \cite{PraPra08}).

Since (as shown above) expanding \KS spacetime can be only of Weyl type II or D in vacuum, Weyl components of boost weight~0 fully determine the Kretschmann scalar, i.e.
\BE
 R_{abcd}R^{abcd}=4({R_{0101}})^2+R_{ijkl}R_{ijkl}+8R_{0j1i}R_{0i1j}-4R_{01ij}R_{01ij} .
\EE
In vacuum $R_{abcd}=C_{abcd}$, and we can use the above compact notation to reexpress  
\BE
R_{abcd}R^{abcd} = 4\WD{} ^2+C_{ijkl}C_{ijkl}+8\WDS{ij} \WDS{ij} -24\WDA{ij} \WDA{ij} .
\label{Kret}
\EE
This will be useful soon in the discussion of singularities.

To conclude, let us observe that Weyl components of boost weight $-1$ and $-2$ contain also derivatives different from $D$, and we cannot study their $r$-dependence at this general level.

\subsubsection{Type D \KS spacetimes: also the second multiple WAND is geodetic}

Type D \KS spacetimes are a special subclass of general \KS metrics. By definition of type D, only boost weight~0 components of the Weyl tensor (as given above) are non-zero in an adapted frame, i.e. they admit a second multiple WAND
not parallel to the \KS (geodetic) vector $\bk$. Here we show that this second WAND is also geodetic (as already known in the special case of Myers-Perry black holes \cite{Hamamotoetal06}, see also \cite{PraPraOrt07}).

First, note that, from (\ref{PhiA}), one has $\WDA{i}{j} =0\Leftrightarrow\omega^2=0$ (cf eqs.~(\ref{s_A}) and  (\ref{Hgeneral2})). In addition, since $\WDS{ij} \neq 0$ (cf~(\ref{Phi_alpha}) and appendix~\ref{app_special}), for $\omega=0$ (i.e., $p=0$) it is easy to see from (\ref{Phi_alpha}) that $\WD{(i)}{(i)} \neq -\Phi$ for all values of $i$. 
Thus, for type D KS spacetimes proposition~6 of \cite{PraPraOrt07} implies that the second multiple WAND is indeed geodetic (since one can always align the frame vector $\bn$ to it, and this is compatible with the assumed condition $\M{i}{j}{0}=0$). This is a general consequence of the Goldberg-Sachs theorem in four dimensions, but it is a non-trivial result for $n>4$ (cf~\cite{PraPraOrt07}). Therefore, it can be viewed as another partial extension of the Goldberg-Sachs theorem, {again} limited to \KS spacetimes. 

\subsection{Singularities} 

The form (\ref{Hgeneral2}) of the \KS function suggests there may be singularities at $r=0$, except in the cases $2p=m$ ($m$ even) and $2p=m-1$ ($m$ odd). In order to invariantly identify possible curvature singularities, let us analyze the behavior of the Kretschmann scalar~(\ref{Kret}). 

Eq.~(\ref{Kret}) clearly consists of a sum of squares, except for the last term, which is negative. For our purposes, it suffices to focus on the first and the last terms determined by (\ref{PhiA}) and (\ref{Phi_alpha}) with (\ref{s_A}) and (\ref{Hgeneral2}).

\subsubsection{`Generic' case ($2p\neq m$, $2p\neq m-1$)} 

Excluding for now the special cases $2p=m$ and $2p=m-1$, from (\ref{Hgeneral2}) one easily finds that for
 $r\sim0$
\be
 \kS\sim r^{-(m-2p-1)} , \qquad D\kS\sim r^{-(m-2p)}, \qquad D^2\kS\sim r^{-(m-2p+1)} .
\ee

Inserting this into~(\ref{PhiA}) and the second of eqs.~(\ref{Phi_alpha}), it is clear that the first term in (\ref{Kret}) will dominate over the last term near $r\sim0$,    
and therefore the Kretschmann scalar will diverge at $r=0$, thus confirming the presence of a curvature singularity. Note that this `generic' case includes, in particular, all non-twisting solutions, for which $p=0$ (and $\WDA{ij} =0$).

As an explicit example, $n>5$ Myers-Perry black holes with only one non-zero spin ($m=n-2$, $p=1$) fall in this subclass, and in appropriate coordinates \cite{MyePer86} one has $2\kS=-\mu r^{5-n}/(r^2+a^2\cos^2\theta)$ ($a$ is the angular momentum parameter). See \cite{MyePer86} for a detailed discussion, including other possible examples with more than one rotation parameter.

\subsubsection{Special cases $2p=m$ and $2p=m-1$}  

Let us now comment on the cases $2p=m$ ($m$ even) and $2p=m-1$ ($m$ odd), for which  $\kS=r\kS_0\prod_{\mu=1}^{m/2}[r^2+(a^0_{(2\mu)})^2]^{-1}$ and $\kS=\kS_0\prod_{\mu=1}^{(m-1)/2}[r^2+(a^0_{(2\mu)})^2]^{-1}$, respectively. 
With these assumptions $\kS$ and its $D$-derivatives are clearly non-singular at $r=0$ (since $p$ denotes the number of non-vanishing terms $a^0_{(2\mu)}$ in (\ref{Hgeneral2})). However, in general, $a^0_{(2\mu)}$ may be functions of spacetime coordinates different from $r$. Given a specific \KS solution, there may thus exist `special points' where some of the $a^0_{(2\mu)}$ vanish.  Let us say, e.g., that $q$ (with $q\ge 1$) of such terms $a^0_{(2\mu)}$ vanish simultaneously there. Then, proceeding as above one finds that at those special points, when $r\to0$ 
\beqn
 & & \kS\sim r^{-2q+1} , \quad D\kS\sim r^{-2q}, \quad D^2\kS\sim r^{-2q-1} \qquad (2p=m) , \\ 
 & & \kS\sim r^{-2q} , \quad D\kS\sim r^{-2q-1}, \quad D^2\kS\sim r^{-2q-2} \qquad (2p=m-1) .
\eeqn  

Again, the first term in (\ref{Kret}) will dominate over the last one and the Kretschmann scalar will diverge at the `special points' when $r\to 0$. 

For example, for both the $n=4$ Kerr solution ($m=2$, $p=1$) and the $n=5$ Myers-Perry metric with only one spin ($m=3$, $p=1$) the only non-zero $a^0_{(2\mu)}$ function is determined by $(a^0_{(2)})^2=a^2\cos^2\theta$. Hence, there is a curvature singularity at $r=0$ and $\theta=\pi/2$ (this is the well-known ring-shaped singularity of the Kerr spacetime). By contrast, the $n=5$ Myers-Perry solution with two non-zero spins (again, $m=3$, $p=1$) contains $(a^0_{(2)})^2=a^2\cos^2\theta+b^2\sin^2\theta$ (in the notation of, e.g., \cite{FroSto03}), which never vanishes. Possible singularities of rotating black hole spacetimes are studied in more generality in \cite{MyePer86}.

\subsection{Special subfamilies}

\label{subsec_special}

We now briefly comment on special subfamilies of expanding \KS solutions, characterized by a non-twisting or a non-shearing \KS vector $\bk$. In relation to the first possibility, let us note that general properties of higher-dimensional non-twisting vacuum spacetimes (not necessarily \KSS) have been recently studied in \cite{PraPra08}, and that all such solutions are explicitly known with the additional assumptions of non-zero expansion and vanishing shear (`Robinson-Trautman spacetimes') \cite{PodOrt06,Ortaggio07}.

\subsubsection{Non-twisting solutions}  

The \KS vector $\bk$ is non-twisting if and only if $L_{ij}$ is a symmetric matrix, i.e. for $p=0$.
This clearly gives 
\be
 L_{ij}=S_{ij}=\frac{1}{r}\mbox{diag}(\underbrace{1,\ldots,1}_{m},\underbrace{0,\ldots,0}_{(n-2-m)}) , 
\ee
so that (cf~(\ref{exp_nondeg})--(\ref{she_nondeg}))
\be
 \theta=\frac{1}{r}\frac{m}{n-2} , \qquad \omega=0 , \qquad \sigma^2=\frac{1}{r^2}\frac{m(n-2-m)}{n-2} .
\ee

From eq.~(\ref{Hgeneral2}), the $r$-dependence of $\kS$ becomes simply 
\be
 \kS=\frac{\kS_0}{r^{m-1}} .
 \label{Hnontwist}
\ee

For Weyl components with boost weight 0 we now have the non-zero components
\beqn
& & \WD{\alpha}{\beta} =\WDS{\alpha}{\beta} =\frac{\kS_0}{r^{m+1}}(m-1)\delta_{\alpha\beta} , \\ 
& & C_{(\alpha)(\beta)(\alpha)(\beta)}=-\frac{2\kS_0}{ r^{m+1}} 
\label{Cijkl_nontwist} .
\eeqn

As mentioned above, all non-twisting \KS solutions contain a curvature singularity at $r=0$. 

Explicit $n$-dimensional examples of non-twisting \KS spacetimes with $m \geq 2$ are static (uniform) black strings and black branes constructed as a direct product of a $(m+2)$-dimensional Schwarzschild black hole with $(n-m-2)$-dimensional flat space (which are of Weyl type D \cite{PraPraOrt07}). In suitable coordinates, these can be written as
\be
 \d s^2=\left[-(2\d r+K\d u)\d u+r^2\d\Omega^2_m+\delta_{AB}\d y^A\d y^B\right]+\frac{\mu}{r^{m-1}}\d u^2 , 
 \label{strings}
\ee 
where the metric in square brackets represents a flat $n$-dimensional spacetime (with $A,B=1,\ldots,n-m-2$, and $\d\Omega^2_m$ being the line element of a $m$-dimensional space of constant curvature $K$), $k_a\d x^a=\d u$ and $2\kS=-\mu r^{-m+1}$. 

The above non-twisting \KS solutions are, in addition, non-shearing iff $m=n-2$ (the Kundt case $m=0$ is ruled out here by the assumption $\theta\neq 0$). In this case, they must belong to the family of higher-dimensional Robinson-Trautman spacetimes \cite{PodOrt06}. In fact, one can see from the results of \cite{PodOrt06} that the only Robinson-Trautman solutions that are also \KS are given by static Schwarzschild-Tangherlini black holes, i.e. by metric~(\ref{strings}) with $m=n-2$ (in which case a cosmological constant can  also be added \cite{PodOrt06} while still preserving the \KS form).\footnote{The $n>4$ Robinson-Trautman family contains, in addition, generalized black holes with non-constant curvature horizons \cite{PodOrt06}, and `exceptional' solutions with `zero mass' $\mu=0$ \cite{PodOrt06,Ortaggio07} (see also \cite{PraPraOrt07}). These cannot be \KS spacetimes because their Weyl components $C_{ijkl}$ contain an $r^{-2}$ term which is absent from~(\ref{Cijkl_nontwist}).}

\subsubsection{Non-shearing solutions ($n$ even)} 

Non-shearing, non-twisting solutions belong to the already discussed \KSS-Robinson-Trautman class, so that we can now focus on non-shearing but twisting solutions. Recall \cite{OrtPraPra07} that $\sigma=0$ and $\omega\neq 0$ can occur only for  even $n$ (and for $\theta\neq 0$). 

We see from (\ref{she_nondeg}) that $\sigma=0$ requires $m=n-2$ (i.e., $L_{ij}$ must be non-degenerate), so that $m$ is also necessarily even. Then, we have to consider two possible situations.
\begin{enumerate}
 \item If $2p=n-2$, $\bk$ is shearfree iff $s_{(2\mu)}=\theta$ for any $\mu=1,2,\ldots,(n-2)/2$ (cf~(\ref{she_nondeg})), so that $a^0_{(2)}=a^0_{(4)}=\ldots a^0_{(n-2)}\equiv a_0$, and  
\be
 \theta=\frac{r}{r^2+a_0^2} , \qquad \omega=\sqrt{n-2}\frac{a_0}{r^2+a_0^2} .
 \label{vvv}
\ee
This agrees with the general behavior found in \cite{OrtPraPra07} (note also that $A_{ik} A_{jk} = \frac{\omega^2}{n-2} \delta_{ij}$, as expected from the optical constraint (\ref{optical_const}) with $\sigma=0$).

The function $\kS$ reduces to
\be
 \kS=\frac{r\kS_0}{(r^2+a_0^2)^{\frac{n-2}{2}}} .
 \label{Hshearfree}
\ee

This includes, for instance, the $n=4$ Kerr solution. Explicit solutions for $n>4$ seem to be presently unknown.
 \item If $2p<n-2$, we have the additional condition $\theta=1/r$. This implies $\omega=a_0=0$, which is the already-discussed non-twisting case.
\end{enumerate}

\section{Conclusions}

\label{sec_conclusions}

We have presented a systematic study of geometric properties of \KS metrics in higher dimensions. We have preliminary discussed general results that do not require any further assumptions. Namely, the \KS vector $\bk$ is geodetic if and only if the energy-momentum tensor satisfies $T_{ab}k^ak^b=0$ (proposition~\ref{prop_geod}). When this happens (e.g., in vacuum) the Weyl tensor is of type II (or more special) and $\bk$ is a multiple WAND (proposition~\ref{prop_II}). Furthermore, we have shown that optical properties of $\bk$ are the same with respect to both the flat background and the full metric. Subsequently, our analysis has focused on vacuum solutions. 

For non-expanding metrics the most general \KS solution is now known, since they are equivalent to the type N Kundt class in vacuum (proposition~\ref{prop_nonexp}). Expanding solutions required a more detailed analysis. Again, they turned out to be algebraically special, but the only possible types are II and D (proposition~\ref{prop_exp}). 
We have also shown that the choice of a possible \KS congruence is restricted in vacuum by an ``optical constraint''. In $n=4$ dimensions this requires $\bk$ to be a shearfree congruence, in agreement with the standard Goldberg-Sachs theorem. For $n>4$ we have proven a partial and apparently weaker extension of this theorem, which however naturally reduces to the familiar result when $n=4$. This extension has been derived here only for \KS solutions but, interestingly, it agrees with} previous results for general vacuum type III/N spacetimes \cite{Pravdaetal04}. Moreover, by integration of the Ricci identities we have fixed the dependence of the optical matrix $L_{ij}$ and of the \KS function $\kS$ on the affine parameter $r$ along $\bk$. This enabled us to prove the presence of a curvature singularity in ``generic'' expanding \KS spacetimes. As a side remark, let us note that we were interested in the $n>4$ case, but our analysis also applies to the $n=4$ case, for which we have rederived various known results previously scattered in several publications (many reviewed, however, in \cite{Stephanibook}).

{In future work, it will be interesting to employ our results to possibly find new expanding \KS solutions, and to understand to what extent the Myers-Perry metrics exhaust the expanding vacuum \KS class. In addition, further investigation will admit non-zero matter fields and {\em generalized} \KS solutions, in which the background is not necessarily flat. These include other important spacetimes such as rotating black holes in (A)dS \cite{HawHunTay99,Gibbonsetal04_jgp} and their NUT extensions \cite{CheLuPop06}.}

\section*{Acknowledgments}
The authors acknowledge support from research plan No AV0Z10190503 and research grant KJB100190702.
M.O. carried out part of his work
 at Departament de F\'{\i}sica Fonamental, Universitat de Barcelona, with a postdoctoral fellowship from Fondazione Angelo Della Riccia (Firenze). 



\renewcommand{\thesection}{\Alph{section}}
\setcounter{section}{0}

\renewcommand{\theequation}{{\thesection}\arabic{equation}}

\section{Frame components of the Riemann tensor ($\bk$ geodetic)}
\setcounter{equation}{0}

\label{app_Riemann}

When $\bk$ is geodesic and affinely parametrized we find the following frame components of the Riemann tensor corresponding to the line element (\ref{Ksmetr})
\beqn
 & & \hspace{-.5cm} R_{0i0j}=0, \qquad R_{010i}=0, \qquad R_{0ijk}=0, \label{R0i0j} \\
 & & \hspace{-.5cm} R_{0101}=D^2\kS, \qquad R_{01ij}=2A_{ji}D\kS+4\kS S_{k[j}A_{i]k}, \label{R0101} \\ 
 & & \hspace{-.5cm} R_{0i1j}=-L_{ij}D\kS-2\kS A_{ki}L_{kj}, \label{R0i1j} \\
 & & \hspace{-.5cm} R_{ijkl}=4\kS(A_{ij}A_{kl}+A_{k[j}A_{i]l}+S_{l[i}S_{j]k}), \label{Rijkl} \\
 & & \hspace{-.5cm} R_{011i}=-\delta_i(D\kS)+2L_{[i1]}D\kS+L_{ji}\delta_j\kS+2\kS(2L_{ji}L_{[1j]}+L_{j1}A_{ji}) , \label{R011i} \\
 & & \hspace{-.5cm} R_{1ijk}=2L_{[j|i}\delta_{|k]}\kS+2A_{jk}\delta_i\kS-2\kS\big(\delta_iA_{kj}+L_{1j}L_{ki}-L_{1k}L_{ji} \nonumber \\
 & & \qquad\quad {}-L_{j1}A_{ki}+L_{k1}A_{ji}+2L_{[1i]}A_{kj}+A_{lj}\M{l}{k}{i}-A_{lk}\M{l}{j}{i}\big) , \label{R1ijk} \\
 & & \hspace{-.3cm} R_{1i1j}=\delta_{(i}(\delta_{j)}\kS)+\M{k}{(i}{j)}\delta_k\kS+(2L_{1j}-L_{j1})\delta_i\kS \nonumber \\
& & \qquad\quad {}+(2L_{1i}-L_{i1})\delta_j\kS+N_{(ij)}D\kS-S_{ij}\Delta\kS+2\kS\big(\delta_{(i|}L_{1|j)} \nonumber \\ 
& & \qquad\quad {}-\Delta S_{ij}-2L_{1(i}L_{j)1}+2L_{1i}L_{1j}-L_{k(i|}N_{k|j)}-2\kS L_{k(i}A_{j)k}\nonumber \\
& & \qquad\quad {}+L_{1k}\M{k}{(i}{j)}-2\kS A_{ik}A_{jk}-L_{k(i}\M{k}{j)}{1}-L_{(i|k}\M{k}{|j)}{1}\big) . \label{R1i1j} 
\eeqn

For certain calculations, it may be useful to note that, using the Ricci identities  (11k, \cite{OrtPraPra07}), the above component $R_{1ijk}$ can also be transformed to the somewhat different form
\beqn 
R_{1ijk}= & & 2A_{jk}\delta_i\kS-L_{ki}\delta_j\kS+L_{ji}\delta_k\kS+2\kS\big(2\delta_{[k}S_{j]i}+2\stackrel{l}{M}_{[jk]}S_{il}\nonumber \\
& & {}-2\stackrel{l}{M}_{i[j}S_{k]l}+2L_{1i}A_{jk}-2L_{1[j}A_{k]i}\big) .
\eeqn

Let us finally emphasise that throughout sections \ref{Sec_nonexp} and \ref{Sec_exp} and in appendix \ref{app_ricci} we restrict to vacuum spacetimes, so that $C_{abcd}=R_{abcd}$ and the Weyl tensor is there given simply by the above expressions.

\section{Expanding solutions with ${D\kS=0}$ do not exist in vacuum}
\setcounter{equation}{0}

\label{app_special}

In the special case when $D\kS=0$, from the vacuum equation (\ref{R01=0}) we get $\omega=0$, i.e. $L_{ij}=S_{ij}$. Then 
eq.~(\ref{vac_ij}) reads
\be
 S_{ik}S_{jk}=(n-2)\theta S_{ij} .
\ee
Using an eigenframe of $S_{ij}$, it is easy to see that the only possible solution has the form 
\be
 S_{ij}=s_{(2)}\,\mbox{diag}(1,0,0,\ldots) ,
 \label{Sij_special}
\ee 
with $s_{(2)}=(n-2)\theta$. 
Now, putting $D\kS=0$, $A_{ij}=0$, $L_{ij}=S_{ij}$ and  eq.~(\ref{Sij_special}) into eqs.~(\ref{R0101})--(\ref{Rijkl}), we find that all components of the Riemann (Weyl) tensor with non-negative boost weight vanish. The algebraic type must thus be III or N. However, by analyzing the Bianchi identities it has been proven in \cite{Pravdaetal04} that the canonical form of the expansion matrix $S_{ij}$ for type N and (non-twisting) type III spacetimes in vacuum is $S_{ij}=s_{(2)}\,\mbox{diag}(1,1,0,0\ldots)$ (cf~eqs.~(50) and (C.20) of \cite{Pravdaetal04}). This is clearly incompatible with (\ref{Sij_special}). Therefore, there do not exist vacuum solutions of the \KS class with $\theta\neq 0$ and $D\kS=0$.

\section{Solving the optical constraint}
\setcounter{equation}{0}

\label{app_constraint}

In this appendix we provide a solution of the optical constraint~(\ref{optical_const}) for the matrix $L_{ij}$, in the case of an expanding \KS vector, i.e. $\theta\neq 0$. We introduce the compact notation
\be
 {\cal F}=\frac{L_{lk}L_{lk}}{(n-2)\theta}=\frac{\sigma^2+\omega^2+(n-2)\theta^2}{(n-2)\theta} ,
 \label{cal_F}
\ee
so that (\ref{optical_const}) simply reads
\be
 L_{il}L_{jl}={\cal F}S_{ij} .
 \label{optical_const2}
\ee

In the following, it will be convenient to analyze separately the two possible cases $\det L\neq 0$ and $\det L=0$.

\subsection{Non-degenerate case}

Let us start with the non-degenerate case $\det L\neq 0$, so that there exists a matrix inverse of $L_{ij}$. 
Let us denote by $L_{ij}^{-1}$ such inverse matrix (or, sometimes, its $(ij)$-element -- there will be no ambiguity according to the context). It is also convenient to define symbols for its antisymmetric and symmetric parts, i.e.
\be
 B_{ij}=L_{[ij]}^{-1} , \qquad C_{ij}=L_{(ij)}^{-1} .
\ee

Now, multiplying (\ref{optical_const2}) by $L^{-1}_{kj}$ 
one gets
\be
 L_{ik}={\cal F}S_{ij}L^{-1}_{kj} ,
 \label{L_nondeg}
\ee
and by further multiplication by $L^{-1}_{li}$ we find $\delta_{lk}={\cal F}C_{lk}$, so that
\be
 L_{ij}^{-1}={\cal F}^{-1}\delta_{ij}+B_{ij}.
 \label{Linv}
\ee 
Eq.~(\ref{L_nondeg}) thus becomes
\be
 L_{ik}=S_{ik}+{\cal F}S_{ij}B_{kj} .
 \label{L_nondeg2}
\ee

The symmetric and anti-symmetric parts of this equation give rise, respectively, to
\beqn
 & & B_{kj}S_{ji}+B_{ij}S_{jk}=0 , \label{symm_L} \\
 & & 2A_{ik}={\cal F}(B_{kj}S_{ji}-B_{ij}S_{jk}) . \label{antisymm_L} 
\eeqn

The symmetric matrix $S_{ij}$ defines a natural frame of orthonormal eigenvectors.
 It is thus convenient to identify our basis vectors $\bm^{(i)}$ with such eigenvectors,\footnote{{\it A priori}, there is no reason to expect that this frame is parallely propagated along $\bk$, a condition that we do not need here. However, we shall see in the following appendix that such an eigenframe is, in fact, compatible with parallel transport.} so that $S_{ij}=\mbox{diag}(s_{(2)},s_{(3)},s_{(4)},\ldots)$.
Eqs.~(\ref{symm_L}) and (\ref{antisymm_L}) then read
\beqn
 & & B_{ki}(s_{(i)}-s_{(k)})=0 , \label{symm_L2} \\
 & & 2A_{ik}={\cal F}B_{ki}(s_{(i)}+s_{(k)}) . \label{antisymm_L2} 
\eeqn

If $B_{ki}=0$ for all values of the indices, then eqs.~(\ref{symm_L2}) and (\ref{antisymm_L2}) are identically satisfied and, by (\ref{Linv}), $L_{ij}={\cal F}\delta_{ij}$, and all eigenvalues of $S_{ij}$ take the same value $s_{(i)}={\cal F}$. 

If, instead, for some components $B_{\bar k\bar i}\neq 0$, eq.~(\ref{symm_L2}) implies that the  corresponding eigenvalues of $S_{ij}$ coincide (i.e., $s_{(\bar i)}=s_{(\bar k)}$). After ordering the basis vectors so as to have multiple eigenvalues next to each other on the diagonal of $S_{ij}$, the matrix $B_{ki}$ must thus be composed of (antisymmetric) blocks with their diagonal on the main diagonal, each block corresponding to a repeated $s_{(\bar i)}$ (e.g., $B_{23}\neq 0$ implies $s_{(3)}=s_{(2)}$ and then, if $s_{(4)}\neq s_{(2)}$, we have $B_{24}=0$, etc.). Moreover, there can be at most one eigenvalue of $S_{ij}$ with multiplicity 1, and this must equal ${\cal F}$ (supposing, e.g., that $s_{(2)}\neq s_{(\hat k)}$ for $\hat k=3,\ldots,n-1$, then one gets $B_{2\hat k}=0$, so that, by (\ref{Linv}), 
$L_{2\hat k}^{-1}=0=L_{\hat k 2}^{-1}$, $L_{22}^{-1}={\cal F}^{-1}$, and thus 
$s_{(2)}={\cal F}$ ; and of course there cannot be distinct eigenvalues both equal to ${\cal F}$). 

Indices $\bar i$, $\bar j$, $\bar k,\ \dots$ take values within a given block (and thus $s_{(\bar i)}=s_{(\bar k)}$), so that eq.~(\ref{antisymm_L2}) now splits into separate equations for each block of the matrix $B_{ki}$, i.e. 
\be
 A_{\bar i\bar k}=s_{(\bar i)}{\cal F}B_{\bar k\bar i} ,
 \label{A_block}
\ee
and components of $A_{ij}$ with indices referring to different blocks necessarily vanish (since thus does $B_{ij}$). The matrix $A_{ij}$ has thus the same block-structure of $B_{ij}$ (or simpler, if some of the $s_{(i)}$ vanish). Eq.~(\ref{A_block}) also implies that all eigenvalues $s_{(i)}$ (for all blocks) must be non-zero, since we are now considering the case $\det L\neq 0$. 

Using (\ref{Linv}), (\ref{L_nondeg2}), and (\ref{A_block}), within each block we can thus write
\be
 L^{-1}_{\bar i\bar j}={\cal F}^{-1}\delta_{\bar i\bar j}+B_{\bar i\bar j} , \qquad L_{\bar i\bar j}=s_{(\bar i)}(\delta_{\bar i\bar j}+{\cal F}B_{\bar j\bar i}) .
\ee
But since these two blocks must be inverse to each other (and since $s_{(\bar i)}\neq 0$ takes the same value for any  $\bar i$ within a block), we necessarily have
\be
 B_{\bar i\bar k}B_{\bar j\bar k}=\frac{{\cal F}-s_{(\bar i)}}{{\cal F}^2s_{(\bar i)}}\delta_{\bar i\bar j} . 
 \label{Bconstraint}
\ee

This implies that any block-matrix $B_{\bar i\bar j}$ must be even-dimensional (and that $({\cal F}-s_{(\bar i)})s_{(\bar i)}>0$), unless it vanishes. When this happens, one has simply $L^{-1}_{\bar i\bar j}={\cal F}^{-1}\delta_{\bar i\bar j}$ and $L_{\bar i\bar j}=s_{(\bar i)}\delta_{\bar i\bar j}$, so that
\be
 s_{(\bar i)}={\cal F} \qquad (\Leftrightarrow B_{\bar i\bar j}=0).
\ee
Hence, if there exist more than one block satisfying $B_{\bar i\bar j}=0$, to such values of  indices
there corresponds only one possible eigenvalue $s_{(\bar i)}={\cal F}$ of the matrix $S_{ij}$, i.e. $S_{ij}$ contains a diagonal block given by ${\cal F}\mbox{diag}(1,1,\ldots,1)$. 

Within each even-dimensional block with $B_{\bar i\bar j}\neq0$, $S_{\bar i\bar j}=s_{(\bar i)}\delta_{\bar i\bar j}$ and  we can thus finally perform a rotation  so as to put this anti-symmetric matrix in a canonical form with two-dimensional anti-symmetric blocks (constrained by (\ref{Bconstraint})) along the diagonal and zeros elsewhere.  

To summarize, we have shown that the matrix $L^{-1}_{ij}$ can always be written in a block-diagonal form. If its antisymmetric part $B_{ij}$ does not vanish identically, there is a number $p\le (n-2)/2$ of two-dimensional blocks of the form
\beqn
 & & {\cal L}^{-1}_{(\mu)}=\left(\begin {array}{cc} {\cal F}^{-1} &  -{\cal F}^{-1}\left(\frac{{\cal F}-s_{(2\mu)}}{s_{(2\mu)}}\right)^{1/2} \\
 {\cal F}^{-1}\left(\frac{{\cal F}-s_{(2\mu)}}{s_{(2\mu)}}\right)^{1/2} & {\cal F}^{-1}
\end {array}
 \right) ,  \label{Linv_block} \\
 & & \mu=1,\ldots, p , \qquad 0\le p\le \frac{n-2}{2}  \nonumber . 
 \eeqn
$s_{(2\mu)}$ need not be all distinct and thus some two-blocks may be identical. 

In addition, there is a $(n-2-2p)\times(n-2-2p)$-dimensional diagonal block
\be
 \tilde{\cal L}^{-1}={\cal F}^{-1}\mbox{diag}(1,1,\ldots,1) .
 \label{Linv_diag}
\ee

Since each block can be inverted separately, finding the explicit form of the matrix $L_{ij}$ is now straightforward. This consists of $p$ blocks of the form 
\be
 {\cal L}_{(\mu)}=\left(\begin {array}{cc} s_{(2\mu)} & \big[s_{(2\mu)}({\cal F}-s_{(2\mu)})\big]^{1/2} \\
 -\big[s_{(2\mu)}({\cal F}-s_{(2\mu)})\big]^{1/2} & s_{(2\mu)} 
\end {array}
 \right) ,
 \label{L0}
\ee
and of 
one $(n-2-2p)\times(n-2-2p)$-dimensional diagonal block
\be
 \tilde{\cal L}={\cal F}\mbox{diag}(\underbrace{1,1,\ldots,1}_{(n-2-2p)}) .
 \label{L0diag}
\ee
More explicitly, the matrix $L_{ij}$ thus takes the form
\beqn
 L_{ij}=\left(\begin {array}{ccccc} \fbox{${\cal L}_{(1)}$} & & & &  \\
 & \fbox{${\cal L}_{(2)}$} & & & \\
 & & \ddots & & \\ 
 & & & \fbox{${\cal L}_{(p)}$} & \label{L_nondeg3} \\
 & & & & \fbox{$\begin {array}{ccc} & & \\ \ \ & \tilde{\cal L} \ \ & \\ & & \end {array}$}
  \end {array}
 \right) . 
\eeqn

\subsection{Degenerate case}

Let us now study the degenerate case $\det L=0=\det L^T$. This implies that there exists a non-zero vector $v$ (living in the `transverse' space spanned by vectors $\bm^{(i)}$) such that $L^Tv=0$. By~(\ref{optical_const2}) this gives also $Lv=0$. If we now choose an orthonormal basis such that, say, $\bm^{(n-1)}$ corresponds to $v$, we can rewrite $L^Tv=0=Lv$ as
\be
 L_{(n-1),i}=0=L_{i,(n-1)} .
\ee
In addition, the optical constraint~(\ref{optical_const2}) now reduces to $L_{\bar i\bar k}L_{\bar j\bar k}={\cal F}S_{\bar i\bar j}$, with the barred indices ranging from 2 to $(n-2)$ only. In other words, we have the same equation in a reduced space. We can then distinguish the two cases $\det L_{\bar i\bar j}\neq 0$ and $\det L_{\bar i\bar j}=0$, and similarly proceed until we find a subspace with a non-degenerate reduced matrix $L$. Finally, we can proceed exactly as in the non-degenerate case. It is thus obvious that the general form of $L_{ij}$ will have the same block diagonal form~(\ref{L_nondeg3}), except that now the diagonal block contains also $(n-2-m)$ zeros,
\be
 \tilde{\cal L}={\cal F}\mbox{diag}(\underbrace{1,\ldots,1}_{(m-2p)},\underbrace{0,\ldots,0}_{(n-2-m)}) ,
 \label{L0diag_deg}
\ee
where $m\equiv\mbox{rank}(L)$, and $0\le 2p\le m$. The non-degenerate case clearly corresponds to $m=n-2$.

\section{Integration of the Ricci identities}
\setcounter{equation}{0}

\label{app_ricci}

The general form of the Ricci identities in higher dimensions has been given in \cite{OrtPraPra07}. Here we consider them only in the case $\theta\neq 0$, and in a form simplified by the fact that $\bl=\bk$ is geodetic and affinely parametrized. Further simplification could be achieved by using a null frame that is parallely transported along $\bk$ (which is always a possible choice \cite{OrtPraPra07}), i.e., $\M{i}{j}{0}=0=N_{i0}$. For certain purposes, however, it is desirable to retain the full freedom of null rotations of $\bn$, and for now we thus only impose
\be
 \M{i}{j}{0}=0 .
 \label{condM}
\ee

We also employ the special form that the Riemann tensor takes in case of \KS spacetimes, as given in appendix~\ref{app_Riemann}, and use the compact notation~(\ref{PhiS})--(\ref{Phi_alpha}) (strictly speaking, this is defined for the Weyl tensor, but in vacuum we can equivalently apply it to the Riemann tensor). Moreover, we keep into account consequences of the vacuum equations discussed in appendix~\ref{app_constraint}.

\subsection{Sachs equations: form of $L_{ij}$}

Under the above assumptions, eq.~(11g,\cite{OrtPraPra07}) becomes
\be
 DL_{ij}=-L_{ik}L_{kj} .
 \label{DLij}
\ee

\subsubsection{Non-degenerate case}

{By methods similar to those of \cite{NP},} the solution to (\ref{DLij}) can be written in terms of its inverse as \cite{OrtPraPra08_2}
\be
 L_{ij}^{-1}=r\delta_{ij}+(L^0)_{ij}^{-1} ,
 \label{inverse_ij}
\ee
where $(L^0)_{ij}^{-1}$ represents $L_{ij}^{-1}$ at $r=0$. For compatibility with the optical constraint, the `initial value' $(L^0)_{ij}^{-1}$ must be of the canonical form obtained in appendix~\ref{app_constraint}. The solution for any $r$ can then be obtained immediately by adding a term $r\delta_{ij}$ to eqs.~(\ref{Linv_block}) and (\ref{Linv_diag}) evaluated at $r=0$.

Hence, the inverse matrix $L^{-1}_{ij}$ consists of $p$ two-blocks and of one $(n-2-2p)\times(n-2-2p)$-dimensional block of the form, respectively,
\beqn
 & & {\cal L}^{-1}_{(\mu)}=\left(\begin {array}{cc} {\cal F}_0^{-1}+r &  -{\cal F}_0^{-1}\left(\frac{{\cal F}_0-s^0_{(2\mu)}}{s^0_{(2\mu)}}\right)^{1/2} \\
 {\cal F}_0^{-1}\left(\frac{{\cal F}_0-s^0_{(2\mu)}}{s^0_{(2\mu)}}\right)^{1/2} & {\cal F}_0^{-1}+r
\end {array}
 \right) , \\
 & & \tilde{\cal L}^{-1}=({\cal F}_0^{-1}+r)\mbox{diag}(1,1,\ldots,1) .
 \label{fff}
\eeqn

Correspondingly, the matrix $L_{ij}$ (\ref{L_nondeg3}) consists of blocks 
\beqn
& &  {\cal L}_{(\mu)}=\frac{1}{1+2rs^0_{(2\mu)}+r^2s^0_{(2\mu)}{\cal F}_0} \nonumber \\
& & \qquad\quad {}\times\left(\begin {array}{cc} s^0_{(2\mu)}(1+r{\cal F}_0) & \big[s^0_{(2\mu)}({\cal F}_0-s^0_{(2\mu)})\big]^{1/2} \\
 -\big[s^0_{(2\mu)}({\cal F}_0-s^0_{(2\mu)})\big]^{1/2} & s^0_{(2\mu)}(1+r{\cal F}_0)
\end {array}
 \right) , \\
& & \tilde{\cal L}=\frac{{\cal F}_0}{1+r{\cal F}_0}\mbox{diag}(1,1,\ldots,1) .
 \label{rrr}
\eeqn 

By construction, this matrix $L_{ij}$ satisfies the optical constraint at ${r=0}$. However, note 
 that the last two equations are of the same form of (\ref{L0}) and (\ref{L0diag}), respectively, provided one rewrites them using 
\be
 s_{(2\mu)}=\frac{s^0_{(2\mu)}(1+r{\cal F}_0)}{1+2rs^0_{(2\mu)}+r^2s^0_{(2\mu)}{\cal F}_0} , \qquad {\cal F}=\frac{{\cal F}_0}{1+r{\cal F}_0} .
\ee
This implies that $L_{ij}$ automatically satisfies the optical constraint for any $r$. In other words, the canonical frame determined by the optical constraint is compatible with the condition $\M{i}{j}{0}=0$ (and, in particular, with parallel transport along $\bk$ if $N_{i0}=0$ is also required).

For practical purposes it will be convenient to simplify the above equations by shifting the affine parameter as $r=\tilde r-1/{\cal F}_0$, and simultaneously defining the new quantities $a^0_{(2\mu)}=\big[s^0_{(2\mu)}({\cal F}_0-s^0_{(2\mu)})\big]^{1/2}/(s^0_{(2\mu)}{\cal F}_0)$.
Then one has (dropping tildes over $r$)
\be
 {\cal L}_{(\mu)}=\frac{1}{r^2+(a^0_{(2\mu)})^2}\left(\begin {array}{cc} r & a^0_{(2\mu)} \\
 -a^0_{(2\mu)} & r
\end {array}
 \right) , \qquad \tilde{\cal L}=\frac{1}{r}\mbox{diag}(1,1,\ldots,1) .
 \label{r_dep_nondeg}
\ee

This general structure of $L_{ij}$ agrees with the results of \cite{PraPraOrt07} (up to reordering the basis vectors) for the specific case of Myers-Perry black holes. 

\subsubsection{Degenerate case}

When $\det L=0$, it is easy to see that the vector $v$ (that satisfies $Lv=0$) can be parallely transported along $\bk$ \cite{OrtPraPra08_2}. We can therefore proceed as in the non-degenerate case to solve the Sachs equations in the `non-zero block' of $L_{ij}$, the remaining part of the matrix keeping its zero form. We thus do not repeat the results here, just replace the last $(n-2-m)$ entries of $\tilde{\cal L}$ by zeros in~(\ref{r_dep_nondeg}).

\subsubsection{Summary}

In the following, we will often need to refer to the elements of the matrices $S_{ij}$ and $A_{ij}$, and we will need a notation that enables us to handle both the non-degenerate and the degenerate case simultaneously. It will  also be convenient to distinguish between indices $\alpha$, $\beta$, \ldots referring to the non-zero eigenvalues of $\tilde{\cal L}$ and $\rho$, $\sigma$, \ldots to its zero eigenvalues. For further compactness, we also use a complex notation (with $i$ being the imaginary unit) and we summarize the non-zero elements of $L_{ij}$ as 
\beqn
  & & s_{(2\mu)}+iA_{2\mu,2\mu+1}=\frac{r+ia^0_{(2\mu)}}{r^2+(a^0_{(2\mu)})^2} , \quad\qquad (\mu=1,\ldots, p) \label{s_A_app} \\
  & & s_{(\alpha)}=\frac{1}{r} \quad\qquad (\alpha=2p+2,\ldots,m+1) , \\
  & & s_{(\rho)}=0 \quad\qquad (\rho=m+2,\ldots,n-1) . \label{s_0_app}
\eeqn

\subsection{Remaining Ricci rotation coefficients}

The above results for $L_{ij}$ enable us to find the $r$ dependence of all Ricci rotation coefficients by integrating the corresponding Ricci identities \cite{OrtPraPra07} that contain a derivative along $\bk$. From now on, however, in addition to~(\ref{condM}) we also use the null rotation freedom on $\bn$ to set
\be
 N_{i0}=0 ,
\ee
i.e. we use a null frame that is {\em parallely transported} along $\bk$.

\subsubsection{Form of $L_{1i}$ and $L_{i1}$}

Let us start from eq.~(11b,\cite{OrtPraPra07}), which now takes the form
\be
 DL_{1i}=-L_{1j}L_{ji} .
 \label{DL1i}
\ee
These equations decouple according to the block structure of $L_{ij}$, and the resulting coefficients can be compactly written in complex form as
\be
 L_{1,2\mu}+iL_{1,2\mu+1}=(l^0_{1,2\mu}+il^0_{1,2\mu+1})(s_{(2\mu)}+iA_{2\mu,2\mu+1}) ,
 \label{L1i}
\ee
where $l^0_{1,2\mu}$ and $l^0_{1,2\mu+1}$ are real integration `constants', and  $s_{(2\mu)}$ and $A_{2\mu,2\mu+1}$ are given in (\ref{s_A_app}).

The solution for coefficients $L_{1i}$ with $i$ corresponding to the diagonal block of $L_{ij}$ is simply 
\be
 L_{1\alpha}=\frac{l_{1\alpha}^0}{r} , \qquad L_{1\rho}=l_{1\rho}^0 .
 \label{L1i_diag}
\ee 

Ricci equations (11e,\cite{OrtPraPra07}), i.e.
\be
 DL_{i1}=-L_{ij}L_{j1} ,
\ee
lead to similar expressions for the coefficients $L_{i1}$, namely 
\beqn
 & & L_{2\mu,1}+iL_{2\mu+1,1}=(l^0_{2\mu,1}+il^0_{2\mu+1,1})(s_{(2\mu)}-iA_{2\mu,2\mu+1}) , \\
 & & L_{\alpha 1}=\frac{l_{\alpha 1}^0}{r} , \qquad L_{\rho 1}=l_{\rho 1}^0 .
 \label{Li1_diag}
\eeqn

\subsubsection{Form of $\M{i}{j}{k}$}

Next, the coefficients $\M{i}{j}{k}$ are analogously determined by eq.~(11n,\cite{OrtPraPra07}), which here takes the form
\be
 D\M{i}{j}{k}=-\M{i}{j}{l}L_{lk} .
\ee
Again, the solutions can be given in natural pairs, each for each block of $L_{ij}$, as distinguished by the value of the last index $k$, i.e. 
\be
 \M{i}{j}{,2\mu}+i\M{i}{j}{,2\mu+1}=(\m{i}{j}{,2\mu}+i\m{i}{j}{,2\mu+1}) (s_{(2\mu)}+iA_{2\mu,2\mu+1}) ,
\ee
with real integration `constants'.
When $k$ corresponds to the diagonal part of $L_{ij}$ we have simply
\be
 \M{i}{j}{\alpha}=\frac{\m{i}{j}{\alpha}}{r} , \qquad \M{i}{j}{\rho}=\m{i}{j}{\rho} . 
\ee
Because of the index symmetries of $\M{i}{j}{k}$ \cite{Pravdaetal04}, we require $\m{i}{j}{k}+\m{j}{i}{k}=0$ for any $i,j,k=2,\ldots,n-1$.

\subsubsection{Form of $\M{i}{j}{1}$}

Eq.~(11m,\cite{OrtPraPra07}) here becomes
\be
 D\M{i}{j}{1}=-\M{i}{j}{k}L_{k1}-2\WDA{ij} . 
\ee 
From (\ref{s_A_app}), one has $DA_{ij}=-2s_{(i)}A_{ij}$ and thus, by (\ref{R0101}), $\WDA{ij} =-D(\kS A_{ij})$. Using also previous results for $\M{i}{j}{k}$ we find
\beqn
 \M{i}{j}{1}= & & \sum_{\mu=1}^p\left[\frac{r}{r^2+(a^0_{(2\mu)})^2}(\m{i}{j,}{2\mu}l_{2\mu, 1}^0+\m{i}{j,}{2\mu+1}l_{2\mu+1, 1})\right. \nonumber \\ 
 & & {}+\left.\frac{a^0_{(2\mu)}}{r^2+(a^0_{(2\mu)})^2}(\m{i}{j,}{2\mu}l_{2\mu+1, 1}^0-\m{i}{j,}{2\mu+1}l_{2\mu, 1})\right] \nonumber \\
 & & {}+\frac{\m{i}{j}{\alpha}l_{\alpha 1}^0}{r}-\m{i}{j}{\rho}l_{\rho 1}^0r+2A_{ij}\kS+\m{i}{j}{1} ,
 \label{Mij1}
\eeqn
where the real `constants' satisfy $\m{i}{j}{1}+\m{j}{i}{1}=0$.

\subsubsection{Form of $L_{11}$}

One can also easily integrate (11a,\cite{OrtPraPra07}), here reduced to
\be
 DL_{11}=-L_{1i}L_{i1}-R_{0101},  
\ee 
to fix the $r$-dependence of $L_{11}$, i.e.
\beqn
 \hspace{-.4cm} L_{11}= & & -D\kS+l_{11}^0+\sum_{\mu=1}^p\big[(l^0_{1,2\mu}l^0_{2\mu,1}+l^0_{1,2\mu+1}l^0_{2\mu+1,1})s_{(2\mu)} \nonumber \\ 
 & & {}+(l^0_{1,2\mu}l^0_{2\mu+1,1}-l^0_{1,2\mu+1}l^0_{2\mu,1})A_{2\mu,2\mu+1}\big] \nonumber \\
 & & {}+\frac{l_{1\alpha}^0l_{\alpha 1}^0}{r}-l_{1\rho}^0l_{\rho 1}^0r .
 \label{L11}
\eeqn

\subsubsection{Form of $N_{ij}$}

Eq.~(11j,\cite{OrtPraPra07}) here reads
\be
 DN_{jk}=-N_{jl}L_{lk}-\WD{kj} .\label{R_j}
\ee
Note that, for $k=2\mu,2\mu+1$, {it is more natural to deal with a complex unknown, namely $N_{j,2\mu}+iN_{j,2\mu+1}$}. Then, using (\ref{PhiS}), (\ref{PhiA}) and (\ref{s_A_app})--(\ref{s_0_app}), eq.~(\ref{R_j}) can be solved obtaining

\beqn
 & & N_{j,2\mu}+iN_{j,2\mu+1}=\left(n_{j,2\mu}^0+in_{j,2\mu+1}^0\right)(s_{(2\mu)}+iA_{2\mu,2\mu+1}) \nonumber \\
& & \qquad\qquad\qquad {}+\kS(s_{(2\mu)}-iA_{2\mu,2\mu+1})(\delta_{j,2\mu}+i\delta_{j,2\mu+1}) , \\
 & & N_{j\alpha}=\frac{1}{r}(n_{j\alpha}^0+\kS\delta_{j\alpha}) , \qquad  N_{j\rho}=n_{j\rho}^0, 
 \label{Nij_diag}
\eeqn
where $n_{ij}^0$ are real `constants'.

\subsubsection{Form of $N_{i1}$ in an adapted parallely transported frame}

So far, the frame vector $\bn$ has not been specified, except for the requirement that it be parallely transported along $\bk$. While retaining the latter condition, we can still perform a null rotation
\BE
 \hbk=\bk, \qquad \hbn =\bn+z_i\bm^{(i)} -\pull z_kz_k\bl , \qquad \hbm{i} =\bm^{(i)} -z_i\bl ,
    \label{nullrot}
\EE
provided $Dz_{i}=0$, so as to simplify some of the Ricci rotation coefficients. 
In particular, if we choose 
\be
 z_{i}=-l^0_{i1} \quad (\mbox{for } i=2,\ldots,m+1) , \qquad z_{\rho}=0,   
\ee
 we obtain (see the transformation properties given in \cite{OrtPraPra07} -- we drop hats from the transformed coefficients) 
\be
 L_{i1}=0  \quad (i=2,\ldots,m+1) ,
\ee
which is equivalent to $l^0_{2\mu,1}=l^0_{2\mu+1,1}=l^0_{\alpha 1}=0$, 
while the $L_{\rho 1}=l^0_{\rho 1}$ are unchanged.  
(Alternatively, one can also set $L_{1i}=0$, but this appears to be less convenient in what follows.) 

With this choice, eq.~(11f,\cite{OrtPraPra07}) becomes
\be
 DN_{i1}=-n^0_{i\rho}l^0_{\rho1}+R_{101i} .
 \label{DNi1}
\ee

In addition, using eq. (\ref{DL1i}) and the commutator (22, \cite{Coleyetal04vsi}), eq.~(\ref{R011i}) reduces to
\beqn
 & & R_{011i}=-D(\delta_i \kS)-2D(L_{1i}\kS) \qquad (i=2,\ldots,m+1), \\
 & & R_{011\rho}=-D(\delta_\rho \kS)-2D(L_{1\rho}\kS)+l^0_{\rho 1}D\kS  .
\eeqn
We can thus straightforwardly integrate eq.~(\ref{DNi1}) and find 
\beqn
 & & N_{i1}=-n^0_{i\sigma}l^0_{\sigma 1}r+\delta_i\kS+2L_{1i}\kS+n_{i1}^0 , \quad (i=2,\ldots,m+1) \\
 & & N_{\rho 1}=-n^0_{\rho\sigma}l^0_{\sigma 1}r+\delta_\rho \kS+(2l^0_{1\rho }-l^0_{\rho 1})\kS+n_{\rho 1}^0 ,
\eeqn
where $\kS$ and $L_{1i}$ are specified by (\ref{Hgeneral2}) and (\ref{L1i}), (\ref{L1i_diag}). 
For certain applications it may be useful to recall that the coefficients $N_{i1}$ vanish if and only if the frame vector $\bn$ is geodetic (an affine parameter corresponding to $L_{11}=0$). 

Note finally that, when $l^0_{2\mu,1}=l^0_{2\mu+1,1}=l^0_{\alpha 1}=0$, eq.~(\ref{Mij1}) simplifies to 
\be
 \M{i}{j}{1}=-\m{i}{j}{\rho}l_{\rho 1}^0r+2A_{ij}\kS+\m{i}{j}{1} ,
\ee
and eq.~(\ref{L11}) becomes
\be
 L_{11}=-D\kS+l_{11}^0-l_{1\rho}^0l_{\rho 1}^0r .
\ee


\providecommand{\href}[2]{#2}\begingroup\raggedright\endgroup

\end{document}